%% file: ms.tex
\documentclass[appendixfloats]{emulateapj}

\shortauthors{Auger et al.}
\shorttitle{Early-type Galaxy Correlations}

\newcommand{\kms}{km\ s$^{-1}$}

\newcommand{\reff}{$r_{\rm e}$}
\newcommand{\mreff}{r_{\rm e}}
\newcommand{\mre}{M$_{r_{\rm e}/2}$}
\newcommand{\mdim}{M$_{\rm dim}$}
\newcommand{\mmre}{{\rm M}_{r_{\rm e}/2}}

\newcommand{\HST}{\emph{HST}}
\newcommand{\be}{\begin{equation}}
\newcommand{\ee}{\end{equation}}
\newcommand{\bea}{\begin{eqnarray}}
\newcommand{\eea}{\end{eqnarray}}
\newcommand{\cet}{c$_{\rm e2}$}
\newcommand{\mcet}{{\rm c}_{\rm e2}}

\begin{document}

\title{The Sloan Lens ACS Survey. X. Stellar, Dynamical, and Total Mass Correlations of Massive Early-type Galaxies}

\author{M. W. Auger\altaffilmark{1,$\dagger$}, T. Treu\altaffilmark{1,2}, A. S. Bolton\altaffilmark{3}, R. Gavazzi\altaffilmark{4}, L. V. E. Koopmans\altaffilmark{5}, P.~J.~Marshall\altaffilmark{1,6}, L. A. Moustakas\altaffilmark{7}, S. Burles\altaffilmark{8}}
\altaffiltext{1}{Department of Physics, University of California, Santa Barbara, CA 93106, USA}
\altaffiltext{2}{Packard Fellow}
\altaffiltext{3}{Department of Physics and Astronomy, University of Utah, Salt Lake City, UT 84112}
\altaffiltext{4}{Institut d'Astrophysique de Paris, UMR7095 CNRS \& Univ. Pierre et Marie Curie, 98bis Bvd Arago, F-75014 Paris, France}
\altaffiltext{5}{Kapteyn Astronomical Institute, University of Groningen, P.O. Box 800, 9700AV Groningen, The Netherlands}
\altaffiltext{6}{Kavli Institute for Particle Astrophysics and Cosmology, Stanford University, Stanford, CA 94305, USA}
\altaffiltext{7}{Jet Propulsion Laboratory, California Institute of Technology, 4800 Oak Grove Drive, M/S 169-237, Pasadena, CA 91109}
\altaffiltext{8}{D. E. Shaw \& Co., L.P., 20400 Stevens Creek Boulevard, Suite 850, Cupertino, CA 95014}
\altaffiltext{$\dagger$} {\texttt mauger@physics.ucsb.edu}

\begin{abstract}
We use stellar masses, surface photometry, strong lensing masses, and
stellar velocity dispersions ($\sigma_{\rm e/2}$) to investigate
empirical correlations for the definitive sample of 73 early-type
galaxies (ETGs) that are strong gravitational lenses from the SLACS survey. The traditional correlations
(Fundamental Plane [FP] and its projections) are consistent with those
found for non-lens galaxies, supporting the thesis that SLACS
lens galaxies are representative of massive ETGs (dimensional mass
M$_{\rm dim}=10^{11}-10^{12}$ M$_\odot$). The addition of
high-precision strong lensing estimates of the total mass allows us to
gain further insights into their internal structure: i) the average
slope of the total mass density profile ($\rho_{\rm tot}\propto
r^{-\gamma'}$) is $\langle \gamma' \rangle = 2.078\pm0.027$ with an
intrinsic scatter of $0.16\pm0.02$; ii) $\gamma'$ correlates with
effective radius (\reff) and central mass density, in the
sense that denser galaxies have steeper profiles; iii) the dark matter
fraction within
\reff/2 is a monotonically increasing function of galaxy mass and size
(due to a mass-dependent central cold dark matter distribution or to baryonic dark matter -- stellar remnants or low mass stars -- if the IMF is non-universal and its normalization increases with mass); iv)
the dimensional mass M$_{\rm dim}\equiv5\mreff\sigma^2_{e/2}/{\rm G}$ is proportional to the total (lensing) mass \mre, and
both increase more rapidly than stellar mass M$_*$ (M$_* \propto \mmre^{0.8}$); v) the Mass Plane
(MP), obtained by replacing surface brightness with surface mass
density in the FP, is found to be tighter and closer to the virial
relation than the FP and the M$_*$P, indicating that the scatter of
those relations is dominated by stellar populations effects; vi) we
construct the Fundamental Hyper-Plane by adding stellar masses
to the MP and find the M$_*$ coefficient to be consistent with zero
and no residual intrinsic scatter. Our results demonstrate that the
dynamical structure of ETGs is not scale invariant and that it is
fully specified by \mre, \reff, and $\sigma_{\rm e/2}$. Although the
basic trends can be explained qualitatively in terms of varying star
formation efficiency as a function of halo mass and as the result of
dry and wet mergers, reproducing quantitatively the observed
correlations and their tightness may be a significant challenge for
galaxy formation models.
\end{abstract}


\section{INTRODUCTION}
\label{sec:intro}

The hierarchical model for structure formation in the context of a
cold dark matter cosmology ($\Lambda$CDM) has been tremendously
successful at describing the large scale features of the Universe
\citep[e.g.,][]{komatsu}. However, there are many important properties
of the Universe at galactic and sub-galactic scales that have evaded a
detailed understanding.  For example, the several bi-modal classes of
galaxies \citep[e.g., red/blue color, early/late-type
morphology;][]{balogh}, the so-called downsizing of star formation
\citep{cowie,cooper,bundy}, the tight empirical correlations between the
observed properties of early-type galaxies
\citep{faber,kormendy,dressler,djorgovski}, correlations (or the lack
of correlations) between these and the local density
\citep{dressler80,cooper}, and the absence of local analogs to
extremely compact high-redshift galaxies
\citep{daddi,trujillo,vandokkum}.

One important step in understanding these phenomena is to explore the
relationship between the baryonic matter that dominates astrophysical
observables and the dark matter that is postulated in the $\Lambda$CDM
model. High precision measurements at galaxy scales are essential to
test whether apparent inconsistencies between the observed universe
and dark matter only cosmological simulations can be reconciled by an
improved understanding of the detailed physical mechanisms governing
baryons and their interaction with DM, or whether a rethinking of the
CDM paradigm might be necessary. Within this context, the origin
of early-type galaxies (ETGs) is currently a point of discord between
observation and theory, and therefore its investigation carries
enormous potential for discovery. Although their formation via merging
of spiral disks is one of the assumptions of the standard paradigm, it
remains to be seen whether this can work in detail.

The tight empirical correlations between the observed properties of
ETGs are a powerful phenomenological tool to relate baryonic and dark
matter. Among these, the correlation between size, surface brightness
and stellar velocity dispersion known as the Fundamental Plane
\citep[][hereafter FP]{faberFP,djorgovski}, and its mass counterpart the
Stellar Mass Plane \citep[M$_*$P; e.g.,][where surface brightness is
replaced by stellar mass]{hydeMP} have provided two key
insights. First, the correlations are ``tilted'' in the sense that
they cannot be explained by assuming that ETGs are a
self-similar family, they obey the virial theorem, and have a constant
mass-to-light ratio \citep{faberFP,ciotti}; mounting evidence, including the alignment of the Mass Plane \citep[MP;][]{b07} with the virial plane, suggests that the dominant cause of the tilt is a mass-dependent central dark matter fraction \citep[e.g.,][]{slacsvii,tortora}. Second, the correlations are remarkably tight \citep{jfk96,hydeMP,graves} implying that for a given size and velocity
dispersion there is very little scatter in the star formation
histories of ETGs. Previous studies have found a slight misalignment between the FP and M$_*$P \citep{hydeMP} which likely results from constructing the FP
with galaxies that span a variety of ages, since age correlates with
both stellar mass (more massive galaxies form earlier) and the stellar
mass-to-light ratio (older stellar populations have larger stellar M/L
for a given stellar mass).

Additional progress in understanding ETGs has been made by attempting to infer the separate
luminous and dark components in the central regions of
galaxies. \citet{tortora} use dynamical masses determined from central
velocity dispersion measurements with stellar masses inferred from
stellar populations synthesis (SPS) models to obtain the central dark
matter fraction; they find a clear dependence on total mass in the
sense that more massive galaxies have larger dark matter fractions
\citep[also see][]{napolitano}, although they assume a mass-traces
light or singular isothermal sphere mass distribution. The assumption
of an isothermal central mass profile seems to be robust, as
illustrated by modeling of X-ray data \citep{humphrey} and strong
lensing and stellar kinematics \citep{slacsiii,k09,barnabe,b10}, although the
origin of this isothermality remains unclear.

We use a sample of early-type gravitational lenses from the Sloan Lens
ACS Survey \citep[SLACS;][]{slacsi,slacsv,slacsix} to investigate the
scaling relations of ETGs, and in particular we
look at the relationships between total and stellar mass with respect
to the other structural parameters of lenses. We exploit multiple
lines of observation -- including strong lensing, stellar dynamics,
and SPS models -- to distinguish between the luminous and dark mass,
and we construct the FP, M$_*$P, and MP for the SLACS ETGs. The paper
is organized as follows. In Section 2 we present the SLACS dataset,
summarize observables listed in previous papers and used for this
analysis, and present new quantities, such as improved estimates for
the total mass within half of the effective radius and the central slope of the
total mass density profile. Section 3 describes bivariate empirical
correlations derived for the SLACS sample, starting from traditional
ones including non-lensing observables and concluding with those
including total mass as derived from strong lensing. Section~4
describes correlations in higher dimensions, including the FP, M$_*$P, MP and
the newly derived Fundamental Hyper-Plane. Section~5 discusses
our results and Section 6 gives a brief summary. A standard cosmological model with $\Omega_m=0.3$, $\Omega_\Lambda=0.7$, and $h=0.7$ is assumed throughout.

\section{The SLACS Survey Data: Observables and Derived Quantities}
\label{S_data}

There are 85 confirmed (grade `A') strong gravitational lenses that have been
discovered by SLACS \citep{slacsv,slacsix}. These include 73 galaxies
with E or S0 morphologies which we take to be the ETGs from SLACS and
which we focus on in this paper. A wealth of data exists for each of
these galaxies, including high-resolution multi-band optical and
near-infrared \emph{Hubble Space Telescope} (\HST) imaging and
fiber-based optical spectroscopy from the Sloan Digital Sky Survey \citep[SDSS;][]{sdss}. These data are used to
infer several fundamental properties of the lenses, which we briefly
detail below and list in Table \ref{T_data}. Some of the observables
and derived quantities have been given in previous papers of the SLACS
series, and will not be repeated in Table~\ref{T_data} for
conciseness. The present compilation is based on the most up-to-date
data and calibrations and supercedes the information presented in
previous papers.

\subsection{Stellar Masses, Luminosities, and Effective Radii}
\label{ssec:sbmass}

The high-resolution multi-band \HST\ imaging is used to infer stellar
masses M$_*$ for each system \citep{slacsix} using SPS models and
assuming either a Chabrier or Salpeter IMF. Four systems only have one
band of \HST\ imaging and we exclude these from
our analysis when stellar masses are required. The SPS models 
provide robust synthetic photometry (sometimes
referred to as $k$-corrections, or k-color corrections) that is approximately
insensitive to the assumed IMF and these models therefore yield
accurate estimates for the $B$- and $V$-band rest-frame luminosity.
The same models also allow us to compute rest frame luminosities of
each galaxy passively evolved to $z=0$ in a self consistent manner;
$B$- and $V$- band luminosities at the redshift of the lens and corrected to $z=0$ can be found in the paper by \citep{slacsix}.
  
Furthermore, we use the \HST\ imaging to determine the effective radius
in each band, and we employ a linear model of effective radius as a
function of wavelength to infer the rest-frame $V$-band effective
radius. We assume $r_{\rm e,\lambda} = a*\lambda + b$ where $a$ and $b$ are determined from a fit to the observations of $r_{\rm e}$ in each filter with rest-frame wavelength given by $\lambda_{\rm c}/(1 + z)$ where $\lambda_{\rm c}$ is the filter central wavelength \citep[e.g.,][]{t01}. Then $r_{\rm e,5500}$  is the effective radius used throughout this paper. As discussed by \citet{slacsvii}, our assumption of \citet{deV}
profiles is valid in the luminosity range covered by the SLACS
sample. The systematic trends in
S{\'e}rsic index $n$ \citep{sersic} for such high-luminosity galaxies are dominated by the intrinsic scatter in the
correlation. As expected, by fitting the SLACS lenses with S{\'e}rsic
models, we find no correlation between $n$ and any of the global
galactic quantities. Likewise the adoption of S{\'e}rsic profiles does not
change any of the trends presented here. Therefore, we limit our
analysis to the simpler and better constrained de Vaucouleurs models.
This does not affect our interpretation of the tilt of the FP and
other key structural parameters, as the effect of varying $n$ is
negligible in the probed mass range \citep{NTB08}.

\subsection{Stellar Velocity Dispersions and Dynamical Masses}
\label{ssec:vdisp}

The SDSS spectroscopy provides an estimate of the luminosity-weighted stellar
velocity dispersion within the $3''$-diameter aperture of the SDSS
fibers which we refer to as $\sigma_{\rm SDSS}$. We use the
prescription of \citet{jorgensen} to infer the velocity dispersion
within half of the effective radius, $\sigma_{e/2}$, for our analysis
of the scaling relations and parameter planes but use the aperture
velocity dispersion in our analysis of the central mass profile
(Section \ref{S_power_law}). 

We combine effective radii and stellar velocity dispersions to
construct a dimensional mass, defined as
\be 
{\rm M_{\rm dim}} \equiv \frac{5 \mreff \sigma^2_{e/2}}{\rm G},
\label{eq:mdyn}
\ee
where the number 5 is the canonical choice for the virial coefficient of dynamical masses of massive ETGs \citep[e.g.,][]{bernardi,cappellari}. We note that the dimensional mass is not actually the dynamical mass \citep[e.g.,][]{slacsvii}, but we choose this form to simplify comparison with dynamical masses.

\subsection{Einstein Radii, Mass Density Profile Slopes,
and Strong Lensing Masses}
\label{ssec:einstein}
Singular isothermal ellipsoid (SIE) lens model fits to the \HST\ data
have been used to derive Einstein radii for each lens systems
\citep{slacsv,slacsix}. The lens and source redshifts are known from
the SDSS spectroscopy, and we can therefore infer the SIE velocity
dispersion, which we refer to as $\sigma_{\rm SIE}$ \citep[e.g.,][]{slacsv}.  Our SIE mass
models also robustly constrain the total projected mass within the
Einstein radii to a precision of a few percent \citep{slacsv}.

We use the available information to constrain power-law total mass
distribution models for each lensing galaxy. The mass distributions
are defined as in \citet{lsd} and \citet{slacsiii,k09}, with $\rho \propto
r^{-\gamma^{\prime}}$, and these models are constrained using the mass
within the Einstein radius, the SDSS stellar velocity dispersion, and
de Vaucouleurs fits to the stellar light distribution. Details of how
the fits are implemented can be found in \citet{suyu} \citep[also see][]{slacsiii,k09}, although for
this analysis we only use our baseline models, characterized by a
\citet{hernquist} model for the stellar distribution and no anisotropy
of the stellar orbits. The isotropy assumption is consistent with
complementary \citep[e.g.,][]{lsd,k09} and more detailed
\citep[e.g.,][]{barnabe} investigations of the SLACS lenses. Note, however, that while mild radial anisotropy would lead to a slightly shallower inferred mass slope \citep[e.g.,][]{k09}, it would cause the inference on the mass within a central aperture to be over- or under-estimated, depending on whether the aperture is smaller or larger than the Einstein radius. We list
the updated mass slopes $\gamma'$ for all lenses with robust kinematic
and lensing data in Table 1; this extends and completes our analysis
of power-law mass models for the SLACS lenses based upon the SDSS spectroscopy
\citep[e.g.,][]{slacsiii,k09}. We use these power-law models to infer the total
projected mass within half of the effective radius (denoted \mre; this radius is chosen because it is well-matched to the typical Einstein radius and therefore leads to the smallest errors from extrapolating the power-law mass model)
which is used in our analysis of the scaling relations and parameter
planes of the lens galaxies.

\subsection{Mass to Light Ratios and Dark Matter Fractions}
\label{ssec:mlfdm}

We construct three different estimators of the central mass to light
ratio, by computing the ratio between total (lensing) mass, dimensional
and stellar mass and luminosity within $r_{\rm e}/2$. The three ratios
are referred to as the total mass-to-light ratio (or \mre/L), the
dimensional mass-to-light ratio (\mdim/L) and the stellar mass-to-light
ratio (M$_*$/L).  When needed, we use the symbol $M/L$ to refer to the
three mass-to-light ratios collectively.

The mass-to-light ratio is relevant for understanding observations but
simulations are more readily understood in the context of dark (or
stellar) mass fraction. We use the stellar masses derived from SPS
models \citep{slacsix} in conjunction with the total mass within half
of the effective radius determined from lensing and dynamics to infer
the {\it projected} (i.e. within a cylinder) dark matter fraction,
$f_{\rm DM} = 1 - {\rm M_*}/\mmre$ where M$_*$ has been scaled to the
stellar mass within $r_{\rm e}/2$ (this is 32\% of the total stellar
mass for a de Vaucouleurs distribution) and any gas is effectively treated as dark matter. We note that the projected DM
fraction is always larger than the (three-dimensional) DM fraction within a sphere of
equal radius because of the contribution
of the outer parts of the halo to the projected quantity. We adopt in
this paper the projected (two-dimensional) dark matter fraction
because it is the most robustly determined quantity and the one
closest to the observables. However, the interested reader can derive
the three-dimensional DM fraction from the projected fraction using the
effective radii and mass density profile slopes given in Table \ref{T_data}.

\begin{deluxetable*}{lccllclll}
\tabletypesize{\scriptsize}
\tablecolumns{9}
\tablewidth{0pc}
\tablecaption{Mass and Structural Parameters for SLACS Early-type Lenses}
\tablehead{
 &
 r$_{\rm e,V}$ &
 $\sigma_{e/2}$ &
 \multicolumn{2}{c}{log[ M$_*$/M$_\odot$ ]} &
  &
 \multicolumn{2}{c}{$f_{\rm DM}$} &
  \\
 \cline{4-5}\cline{7-8}
 \colhead{Lens Name} &
 \colhead{(kpc)} &
 \colhead{(km~s$^{-1}$)} &
 \colhead{Chab} &
 \colhead{Salp} &
 \colhead{log[ \mre/M$_\odot$ ]} &
 \colhead{Chab} &
 \colhead{Salp} &
 \colhead{$\gamma^{\prime}$}
}
\startdata
\input{T_lens_data.tex}
\enddata
\label{T_data}
\tablecomments{Columns are: 1) Lens name; 2) effective radius,
corrected to the rest-frame V-band, in kpc; 3) central velocity
dispersion within half of the effective radius; 4-5) total stellar
mass assuming a Chabrier (col. 4) or Salpeter (col 5) IMF; 6) total
mass within half of the effective radius, as determined from our
power-law mass distribution models; 7-8) dark matter fraction within
half of the effective radius for Chabrier and Salpeter IMFs; and 9)
slope of the power-law mass distribution, $\rho \propto r^{-\gamma}$.}
\end{deluxetable*}

\section{Bivariate correlations}
\label{sec:2D}

Bivariate empirical correlations between properties of ETGs, e.g.,
the L~-~$\sigma$ \citep{faber} and L~-~\reff\ \citep{kormendy}
relations, are extremely useful for a number of practical
applications, including modeling of complex data and simulations of
mock catalogs. Although ETGs are known to be at least a
two parameter family, the bivariate correlations encode critical
information about the distribution of ETGs in higher-dimensional
parameter spaces and therefore can be used to provide additional
tests of theoretical models. For example, although dry mergers generally 
move ETGs inside the Fundamental Plane, they also tend to move galaxies
away from its two dimensional projections \citep{nipoti03,boylankolchin}.

We fit several bivariate empirical correlations to the SLACS data described in Section 2. We have determined linear fits to each correlation (these correlations are frequently between the logarithm of physical quantities and a linear fit therefore represents a power-law model) that account for the errors in both the
dependent and independent variables as well as covariance between the measurement errors, and we also explicitly determine
the intrinsic scatter. We use a Python implementation of the fitting
technique proposed by \citet{kelly}, which uses a Bayesian framework
to avoid biases introduced by inappropriate choices for the prior
distributions of the independent variables; we have found that this is
particularly important when the errors on the independent variables
are significantly larger than the errors on the dependent variables.

We have restricted our analysis of these correlations to the early-type lenses, which we assume share similar formation and evolution histories. However, \citet{jiang} suggest that the SLACS lenses are either not an homologous population or the stellar velocity dispersions have significantly under-estimated systematic errors \citep[e.g.,][who show that different velocity dispersion codes produce systematically different results]{hyde}. We find that 6 galaxies in our sample are significant outliers of the hyper-plane relation between size, velocity dispersion, stellar mass, and total mass given by Equation \ref{E_hyperplane} (these systems are indicated in Table \ref{T_data}). We have investigated the nature of these outliers but find that they generally do not stand out from the other lensing galaxies; none of these discrepant objects have disky structure, nor do they have anomalous spectral features. Furthermore, we have determined the stellar velocity dispersions for each object using three independent codes but the codes do not find a significant difference between these six objects and the others. Nevertheless, we use an abundance of caution and exclude these six objects from the fits that include intrinsic scatter, as these objects tend to dominate those relations.


We begin with ``traditional'' correlations between non-lensing
observables in \S~\ref{ssec:trad2D}. The main purpose of this section
is to compare our inferred correlations with those inferred from samples
of non-lens galaxies to test the hypothesis that SLACS lens galaxies
are representative of the overall population of massive ETG. Previous
SLACS papers have investigated this issue based on a number of tests
and have found no evidence for any difference between the SLACS lenses and
ETGs with similar velocity dispersions
\citep{slacsii,slacsv,slacsviii,slacsix}. This study updates and
extends some of those tests by considering the larger
sample and including in the analysis relations based on stellar mass.

We introduce lensing observables in \S~{\ref{S_mass_correlations} to
investigate correlations between stellar, total, and dimensional masses,
as a means to constrain the virial coefficient, the initial mass
function, and dark matter content of ETGs. In~\S~\ref{S_power_law} we study the distribution of slopes of
the total mass density profile $\gamma^{\prime}$, extending the
analysis previously published by \cite{k09}. The goal of this analysis
is twofold. From a galaxy formation point of view, the distribution of
total mass density profiles constrains the relative distribution of
baryons and dark matter and therefore constrains quantities such as
the star formation efficiency. From the point of view of gravitational
lensing studies, the distribution of $\gamma^{\prime}$ is an essential
piece of information for inferences regarding, e.g., cosmological
parameters from gravitational time delays and lens statistics \citep[e.g.,][]{suyu,dobke,oguri}.

Finally, in \S~\ref{ssec:mlfdm2d} we examine variations in central
mass-to-light ratio and dark matter fraction with galaxy global
properties again as a means to investigate the initial mass function
and dark matter content of ETGs.

\subsection{Traditional (non-lensing) correlations}
\label{ssec:trad2D}

The $\sigma_{e/2}$-M$_*$ relation is shown in
Figure~\ref{F_sigma_mstar} for a Chabrier IMF. The relation is
slightly shallower than for SDSS galaxies in general ($a =
0.18\pm0.03, b = 2.34\pm0.01$ with $\sigma_{e/2}$ in units of km
s$^{-1}$ and M$_*$ in units of $10^{11} {\rm M_\odot}$), largely due
to explicitly including intrinsic scatter in the relation (Table \ref{T_traditional}). Nevertheless, the \citet{hyde} relations are acceptable fits to the SLACS
data; this is another indication that SLACS lenses constitute a
velocity dispersion-selected subsample of the general population of
massive ETGs \citep{slacsv}.

\begin{figure}
\begin{center}
 \includegraphics[width=0.48\textwidth,clip]{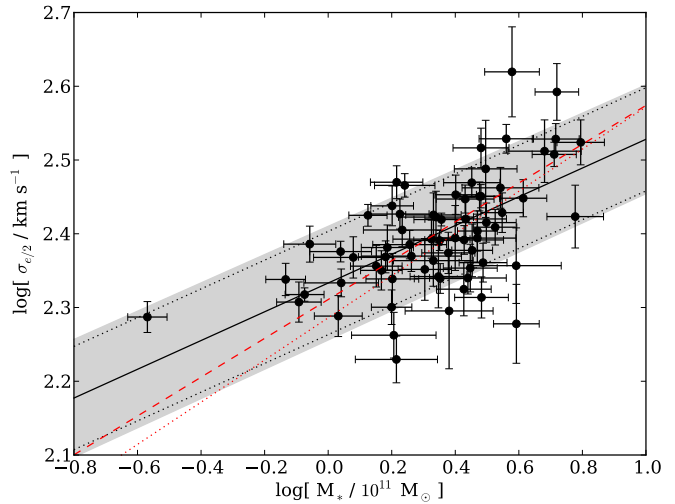}
\end{center}
 \caption{The $\sigma_{e/2}$-M$_*$ relation for SLACS lenses. The
 solid black line is a linear fit to the relation (including scatter), the dotted black
 lines indicate the intrinsic scatter, the gray band indicates the
 quadrature-sum of the scatter and the uncertainty on the linear fit,
 the red dotted line is the linear fit from \citet{hyde}, and the red
 dashed line is their quadratic fit. We find that the SLACS
 $\sigma_{e/2}$-M$_*$ relation is consistent with the SDSS relations,
 although our formal fit is shallower due to our explicit treatment of
 intrinsic scatter.}
 \label{F_sigma_mstar}
\end{figure}

\begin{deluxetable}{lllll}
\tabletypesize{\scriptsize}
\tablecolumns{5}
\tablewidth{0pc}
\tablecaption{$r_{\rm e} - {\rm M_*}$ and $\sigma_{e/2} - {\rm M_*}$ Relations}
\tablehead{
 \colhead{Y} &
 \colhead{X} &
 \colhead{Slope $a$} &
 \colhead{Intercept $b$} &
 \colhead{Scatter}
}
\startdata
$\sigma_{e/2}$  &  M$_*$  &  $0.24\pm0.02$  &  $2.32\pm0.01$ &  \nodata  \\
$\sigma_{e/2}$  &  M$_*$  &  $0.18\pm0.03$  &  $2.34\pm0.01$ &  $0.04\pm0.01$  \\
$r_{\rm e}$  &  M$_*$  &  $0.89\pm0.04$  &  $0.52\pm0.02$ &  \nodata  \\
$r_{\rm e}$  &  M$_*$  &  $0.81\pm0.05$  &  $0.53\pm0.02$ &  $0.05\pm0.02$  \\
\enddata
\label{T_traditional}
\tablecomments{Fits are of the form log Y = a log X + b with M$_*$ in units of $10^{11}~{\rm M}_\odot$, $\sigma_{e/2}$ in units of km~s$^{-1}$, and $r_{\rm e}$ in units of kpc. Each fit is performed twice, either including intrinsic scatter or assuming zero intrinsic scatter. Note that the inclusion of intrinsic scatter has a significant affect on both the size-mass and velocity dispersion-mass relations. These fits are for a Chabrier IMF, but the slope is unaltered assuming a Salpeter IMF.}
\end{deluxetable}

We examine the \reff-M$_*$ relation in Figure~\ref{F_re_mstar}. We find that the SLACS lenses have a somewhat
steeper relation ($a = 0.81\pm0.05, b = 0.53\pm0.02$ for a Chabrier
IMF with \reff, measured in kpc and M$_*$ in units of $10^{11}$
M$_\odot$; see Table \ref{T_traditional}) than non-lensing galaxies in SDSS \citep{shen,hyde}. This is due in part to the curvature in the \reff-M$_*$ relation;
SLACS is dominated by galaxies with M$_* > 10^{11}{\rm M_\odot}$
whereas $10^{11}{\rm M_\odot}$ is the mid-point of the data fit in the
SDSS samples. Indeed, we see in Figure \ref{F_re_mstar} that the SLACS
lenses follow the high-mass end of the quadratic fit of \citet{hyde}
reasonably well. Additionally, SLACS is effectively a sample selected
on velocity dispersion and the typical velocity dispersion at fixed
effective radius is likely larger than non-lensing SDSS galaxies; this
would lead to higher inferred stellar masses at fixed effective radius
(see Figure \ref{F_sigma_mstar}), as is seen in our data. Furthermore, \citet{tortora} find a slope of $0.73\pm0.12$ for the \reff-M$_*$ relation of massive (M$_* > 10^{11.1}$ M$_\odot$ local ETGs, completely consistent with our results.

There are two components of the velocity dispersion selection function in the
SLACS sample. The first comes from the lensing cross section which
scales approximately with $\sigma^4$. The second one comes from the
selection function of the SDSS survey. Firstly, SDSS is a flux-limited
sample so that high luminosity, and therefore high $\sigma$, galaxies
are over-represented because they are visible over a larger
volume \citep{hyde}. Secondly, SDSS has finite resolution and thus
ultra-compact galaxies are difficult to identify \citep[e.g.,][]{trujillo09,taylor,stockton}.

We can correct for the lensing bias to make the SLACS sample directly comparable to the parent SDSS sample by weighting each galaxy's contribution to the posterior distribution function by an exponent proportional to $\sigma^{-4}$ (in $\chi^2$ terms this would be equivalent to weighting each galaxy's contribution to the $\chi^2$ by the same factor). Additionally, we can weight each galaxy by the volume in which it could be observed to provide a more direct comparison with \citet{hyde}. We find that these weighting schemes alter the fits that do not include intrinsic scatter significantly more than the fits with intrinsic scatter; the fits with intrinsic scatter yield consistent results with and without weighting, and we therefore quote the unweighted fits throughout.

\begin{figure}
\begin{center}
 \includegraphics[width=0.48\textwidth,clip]{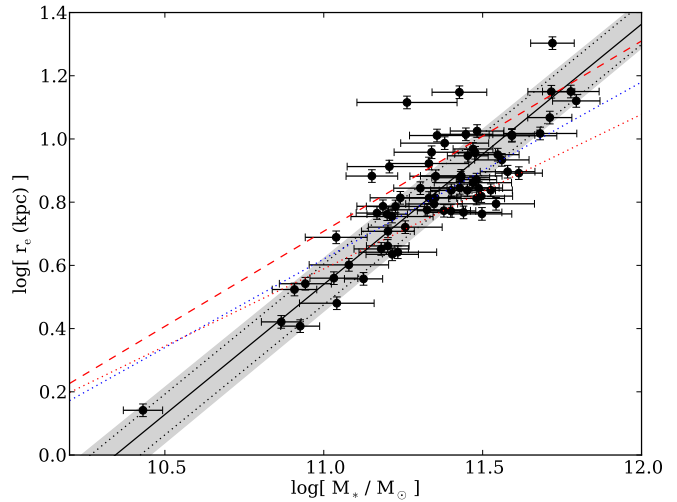}
\end{center}
 \caption{The \reff-M$_*$ relation for SLACS lenses. The solid black
 line is a linear fit to the relation; the dotted black lines indicate
 the intrinsic scatter and the gray band indicates the quadrature-sum
 of the scatter and the uncertainty on the linear fit. The blue dotted
 line is the relation fit by \citet{shen} while the red dotted line is
 the linear fit from \citet{hyde} and the red dashed line is their
 quadratic fit.}
 \label{F_re_mstar}
\end{figure}

\subsection{Correlations between Stellar, Dynamical, and Total Mass}
\label{S_mass_correlations}

Bivariate correlations between mass estimators are shown in
Figure~\ref{F_mass_comparison} and the parameters of linear fits to these relations are
given in Table~\ref{T_mass_comparison}. The linear correlation between
dimensional mass \mdim\ and lensing (or total) mass \mre\ indicates that
the virial coefficient is constant over the range in mass probed by
the SLACS sample, in agreement with our previous result
\citep{slacsvii}. The average value of 
the dimensionless parameter akin to the virial coefficient, log \cet, defined by
$$
{\rm log}\ \mmre = {\rm log}\ \frac{\mcet \mreff \sigma^2_{e/2}}{2 G},
$$
is found to be $0.53\pm0.09$ and the scatter is $0.06\pm0.01$ dex; both of these are consistent with our previous measurement \citep{slacsvii}.
The uniformity of the
virial coefficient (i.e., the very small intrinsic scatter) does not imply exact scale-invariance of the
mass-dynamical structure of early type galaxies. In fact, as shown by
\citet{NTB08}, the observed virial coefficient can be reproduced
by a variety of two component mass models with a broad distribution
of central dark matter fractions. The uniformity of the virial
coefficient, however, restricts the possible range of acceptable models
for ETGs; in particular, models with extreme
orbital anisotropies or that depart significantly from an isothermal
total mass density profile are ruled out \citep{NTB08}.

\begin{figure*}[ht]
\begin{center}
 \includegraphics[width=\textwidth,clip]{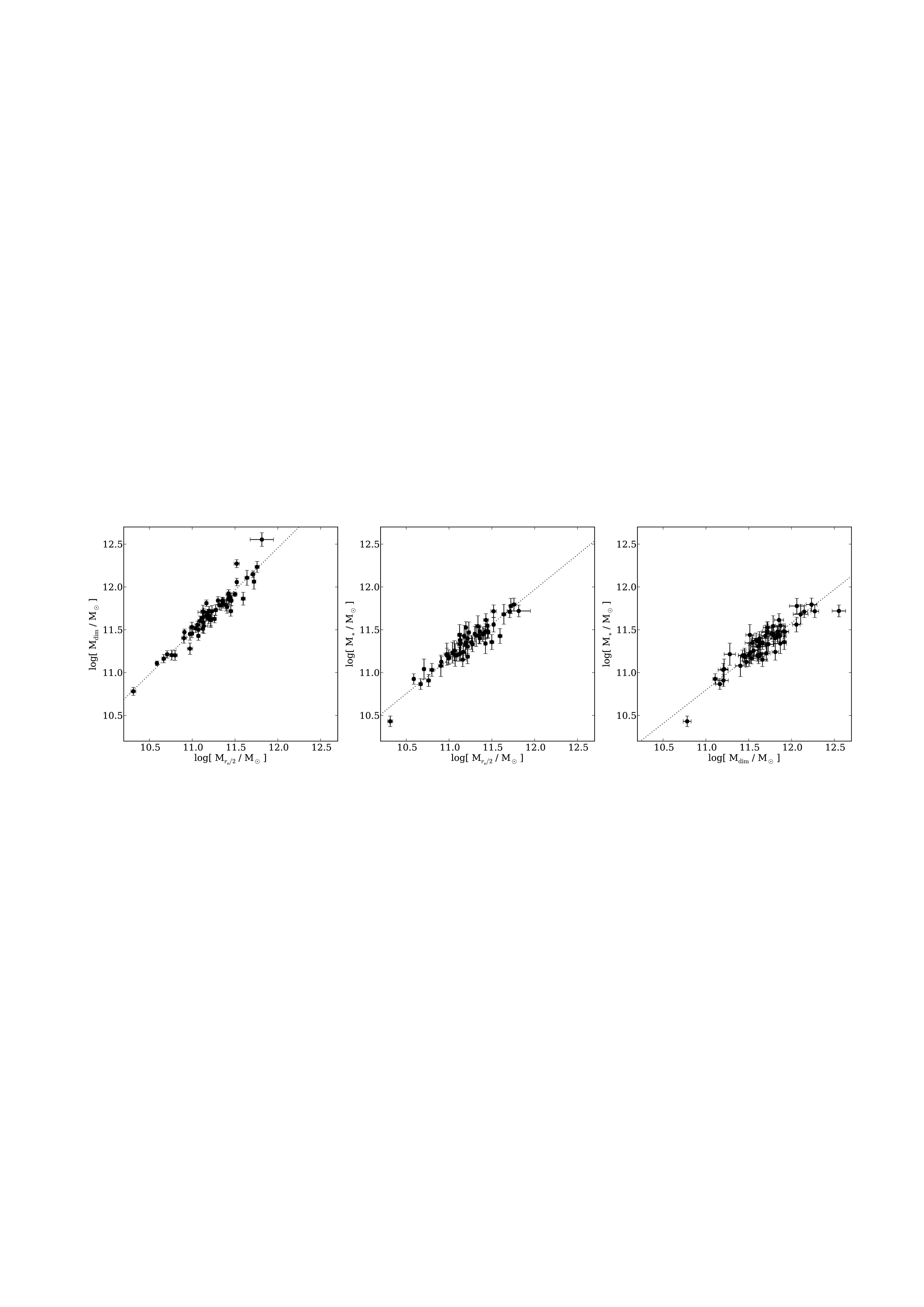}
\end{center}
\caption{Bivariate correlations between dimensional (M$_{\rm dim}$),
total (\mre) and stellar mass (M$_*$). The best fitting linear
relations are shown as dotted lines. Their coefficients are given in
Table~\ref{T_mass_comparison}. Note the linear relation between
dimensional and total mass, and the non-linearity of the other two
relations. These are consistent with a constant virial coefficient and
an increase with mass of the dark matter content and/or stellar
initial mass function normalization.}
\label{F_mass_comparison}
\end{figure*}

The correlation between total mass and stellar mass has the same
amount of intrinsic scatter as the one with dimensional mass, once the
larger errors associated with stellar mass are accounted for. However,
the slope of the correlation differs significantly from unity. This is
consistent with the well-known ``tilted'' slope of the correlation
between dynamical and stellar mass \citep{faberFP,cappellari,gallazzi,bundy07,rettura,graves}
and can be explained in terms of a
varying dark matter fraction and/or stellar IMF normalization with
mass, in the sense that more massive galaxies have a higher fraction of (baryonic or otherwise) dark matter. Interestingly,
however, the small intrinsic scatter indicates that {\it at fixed
mass} the dark matter fraction and/or stellar IMF normalization are
tightly constrained. We will return to these points in
\S~\ref{ssec:mlfdm2d}.

\begin{deluxetable}{lllll}
\tabletypesize{\scriptsize}
\tablecolumns{5}
\tablewidth{0pc}
\tablecaption{Correlations Between Masses}
\tablehead{
 \colhead{Y} &
 \colhead{X} &
 \colhead{Slope} &
 \colhead{Intercept} &
 \colhead{Scatter}
}
\startdata
M$_{\rm dim}$ & M$_{r_{\rm e}/2}$  &  $0.97\pm0.02$  &  $\phantom{-}0.51\pm0.03$  & \nodata \\
M$_{\rm dim}$ & M$_{r_{\rm e}/2}$  &  $0.98\pm0.03$  &  $\phantom{-}0.50\pm0.04$  & $0.06\pm0.01$ \\
M$_*$         & M$_{r_{\rm e}/2}$  &  $0.81\pm0.03$  &  $\phantom{-}0.35\pm0.04$  &  \nodata \\
M$_*$         & M$_{r_{\rm e}/2}$  &  $0.80\pm0.04$  &  $\phantom{-}0.36\pm0.05$  & $0.03\pm0.02$ \\
M$_*$         & M$_{\rm dim}$      &  $0.80\pm0.04$  &  $-0.01\pm0.06$ &  \nodata \\
M$_*$         & M$_{\rm dim}$      &  $0.79\pm0.04$  &  $\phantom{-}0.01\pm0.08$  &  $0.04\pm0.02$
\\
\enddata
\label{T_mass_comparison}
\tablecomments{Correlation between masses. All fits are done in
logarithmic scales and using masses in units of 10$^{10}$ M$_{\odot}$,
to reduce covariance. Fits without and with intrinsic scatter are
given for each pairwise combination. For example, the first line
contains the results of fitting $\log$M$_{\rm dim}/10^{10}$M$_\odot =
a \log$M$_{r_{\rm e}/2}/10^{10}$M$_\odot + b$, where $a$ and $b$ are the
slope and intercept, respectively. The second line adds an additional
Gaussian component with average zero to represent intrinsic
scatter. The width of the Gaussian intrinsic scatter is $0.06\pm0.01$
dex }
\end{deluxetable}

\subsection{Correlations with the Slope of the Total Mass Density Profile}
\label{S_power_law}

We first examine the overall distribution of slopes of the mass
density profile in a joint-framework wherein the $\gamma'$ of all of
the early-type lenses are assumed to be drawn from a normal
distribution parameterized by an average $\gamma^{\prime}_0$ and a
dispersion $\sigma_\gamma^{\prime}$. We find $\gamma^{\prime}_0 =
2.078\pm0.027$ and $\sigma_\gamma^{\prime} = 0.16\pm0.02$, consistent with and extending the results of
\citet{k09}. The SLACS early-type lenses appear to be slightly
super-isothermal (i.e., the density profiles are typically steeper
than isothermal by $\sim5\%$) with an intrinsic spread of
$\approx10\%$. We note, however, that we have assumed that the galaxies are
isotropic; as discussed in \citet{k09}, a modest amount of
radial anisotropy \citep[$\beta \lesssim 0.5$, consistent with][]{gerhard} is sufficient to produce an isothermal slope, $\gamma^{\prime}_0 = 2$. We cannot directly probe the anisotropy with our data \citep[but see][which provides an indirect constraint]{k09} and we therefore impose the isotropic model.

It was previously found that the ratio of the observed stellar
velocity dispersion and the SIE model velocity dispersion, $f_{\rm
SIE} \equiv \sigma_{e/2}/\sigma_{\rm SIE}$, is strongly correlated
with the logarithmic density slope \citep{slacsviii}. Our updated
analysis confirms this result (Figure \ref{F_fsie}) and the best-fit
linear relation, taking into account the covariance between $f_{\rm
SIE}$ and $\gamma^{\prime}$, is given by $\gamma^{\prime}-2 =
(2.67\pm0.15)(f_{\rm SIE} - 1) + (0.20\pm0.01)$. Note that this trend is a
natural consequence of how $\gamma^{\prime}$ is determined and does not provide any significant physical insights; instead, it provides a useful shortcut for determining the power-law slope from $\sigma_{e/2}$ and $\sigma_{\rm SIE}$ without needing to perform Jeans modeling.

\citet{k09} found that the power-law slope did not correlate strongly
with many of the global galaxy observables, including redshift, the
ratio of the Einstein and effective radii, the central lensing mass,
and $\sigma_{\rm SIE}$. This updated analysis does not substantially
change these results since we are only introducing 20\% more
systems. However, we now also investigate correlations with $r_{\rm
e}$, $\sigma_{e/2}$, and the central surface mass density $\Sigma_{\rm
tot} \equiv \mmre/r^2_{\rm e}$. We find non-negligible correlations
(non-zero slopes with greater than 3-$\sigma$ significance) with \reff\ and
$\Sigma_{\rm tot}$ but no clear trend with $\sigma_{e/2}$
(Figure \ref{F_gammap_correlations}); the
correlation with $\Sigma_{\rm tot}$ is the tightest and most significant (Table
\ref{T_gammap_correlations}). This is expected,
since a steeper mass density profile implies a higher central surface
mass density, and explains at least in part the intrinsic dispersion
in the average mass density profile. However, the residual intrinsic
dispersion (0.12, i.e. 6\% in slope) indicates that there may be
additional observables that correlate with the inferred slope. One
such parameter could be local environment, due to tidal effects on the
outer halos or to contamination to the lensing convergence by external
mass along the line of sight. The former has been tentatively detected
using the SLACS sample at marginal levels of significance \citep{auger,slacsviii}, and it has been observed in clusters \citep[e.g.,][]{natarajan}.
The latter does not seem to be significant in the SLACS
sample, as inferred from the minimal level of misalignment between the
major axis of the light and mass. Another element that may contribute to the scatter is anisotropy of the stellar orbits. However, higher precision measurements will be required to determine whether the residual scatter is stochastic in
nature and/or if there are residual and undetected small systematic
trends.

\begin{figure}
\begin{center}
 \includegraphics[width=0.48\textwidth,clip]{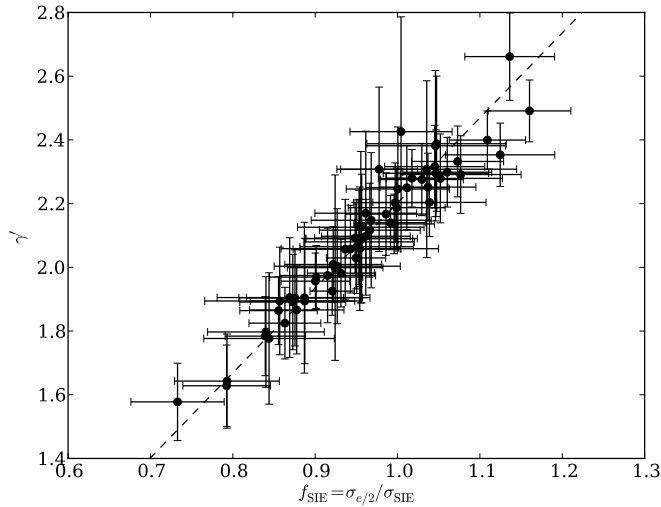}
 \caption{The relationship between the ratio of the observed stellar
 velocity dispersion and the SIE model velocity dispersion, $f_{\rm
 SIE} \equiv \sigma_{e/2}/\sigma_{\rm SIE}$, and the logarithmic
 density slope $\gamma^{\prime}$. The correlation is very tight, with
 evidence for little intrinsic scatter ($\sigma_{int} = 0.02$). Note
 that although the errors on $f$ and $\gamma^{\prime}$ are strongly
 correlated our analysis takes this correlation into account and finds
 a significant excess correlation. The tightness of the relation implies
that $f_{\rm SIE}$ can be a useful proxy for $\gamma^{\prime}$ without
performing the joint lensing and dynamics modeling.}
 \label{F_fsie}
\end{center}
\end{figure}

\begin{figure}
\begin{center}
 \includegraphics[width=0.48\textwidth,clip]{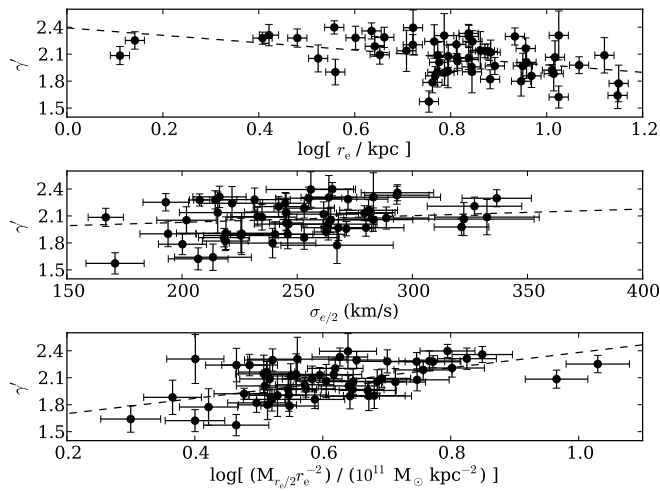}
 \caption{Correlations between $r_{\rm e}$, $\sigma_{e/2}$, and
 $\Sigma_{\rm tot}$ (see Table \ref{T_gammap_correlations} for the
 linear fits). The correlation with $\sigma_{e/2}$ is not significant although the \reff\ and $\Sigma_{\rm tot}$ correlations are found at greater than 3-$\sigma$ significance. All correlations show significant scatter, and the effective radius correlation is somewhat weaker than the central surface mass density
correlation. This latter correlation is expected;
 steeper power-law slopes imply higher central concentrations, and
 higher central concentrations imply increased central surface mass
 densities.}
 \label{F_gammap_correlations}
\end{center}
\end{figure}

\begin{deluxetable}{llll}
\tabletypesize{\scriptsize}
\tablecolumns{4}
\tablewidth{0pc}
\tablecaption{Correlations with $\gamma^{\prime}-2$}
\tablehead{
 \colhead{X} &
 \colhead{Slope} &
 \colhead{Intercept} &
 \colhead{Scatter}
}
\startdata
log $r_{\rm e}$  &  $-0.41\pm0.12$  &  $\phantom{-}0.39\pm0.10$  &  $0.14\pm0.02$ \\
$\sigma_{e/2}$  &  $\phantom{-}0.07\pm0.08$  &  $-0.12\pm0.21$  &  $0.17\pm0.02$ \\
$\Sigma_{\rm tot}$  &  $\phantom{-}0.85\pm0.19$  &  $-0.47\pm0.12$  &  $0.12\pm0.02$ \\
\enddata
\label{T_gammap_correlations}
\tablecomments{$r_{\rm e}$ is in units of kpc, $\sigma_{e/2}$ is in units of 100~\kms, and $\Sigma_{\rm tot}$ is in units of $10^{11}$~M$_\odot$~kpc$^{-2}$.}
\end{deluxetable}

There is tentative evidence for a slight anti-correlation between $\gamma^{\prime}$ and the total and stellar masses, as one might infer from the trends with radius and surface mass density, although neither of these anti-correlations between slope and mass are statistically significant given our sample size and data quality. Nevertheless we note that an anti-correlation between mass and density slope can arise as a result of mergers \citep[e.g.,][]{bk04}, although analytic models of halo collapse may predict a \emph{positive} correlation between the mass and central density slope of the dark matter halo \citep[e.g.,][]{delpopolo}.

We have previously combined the SLACS lenses with a higher-redshift
set of lenses from the Lensing Structure and Dynamics
\citep[LSD;][]{lsd} sample and found marginal evidence that the
central mass-density slope evolves with redshift
\citep{slacsiii}. However, the relationship between $\gamma^{\prime}$
and $\Sigma_{\rm tot}$ was not explicitly accounted for in that
analysis; a full investigation of the evolution of $\gamma^{\prime}$
including this effect will require an expanded sample of high-redshift
lenses.

\subsection{Central mass-to-light ratios and dark matter fraction correlations}
\label{ssec:mlfdm2d}

The SLACS dataset presents a unique opportunity to investigate the
central mass-to-light ratio and dark matter fraction in galaxies
beyond the local universe. We first look at the relationship between M/L and six other
parameters: $B$-band luminosity at $z = 0$, $V$-band luminosity at $z
= 0$, $\sigma_{e/2}$, M$_{\rm dim}$, M$_*$, and
\mre. These trends are shown in Figure \ref{F_ml_fits} and listed in
Table \ref{T_ml_fits}. It is clear that there is a significant trend
between all of these parameters and total M/L, while M$_*$/L is, at the 2-$\sigma$ level, independent of the quantities investigated, and we also note that M$_*$/L and the total M/L correlate more strongly with $\sigma_{e/2}$ than with mass or luminosity. These results are in excellent agreement with \citet{cappellari} and \citet{tortora}, who investigate the
stellar M/L and dynamical M/L for samples of E and S0
galaxies, and with results from the FP and M$_*$P of SDSS galaxies
\citep{hydeMP,graves}. However, \citet{grillo10} find a somewhat steeper trend between M$_*$ and M$_*$/L$_{B}$ for the SLACS lenses ($a = 0.18$), which we attribute to differences in the stellar mass determination due to assumptions about age and metallicity \citep[e.g.,][]{slacsix}. For very massive ETGs like the 
ones in the SLACS sample, the differences in stellar population
properties are not sufficiently large to account for large changes in the
stellar M/L with a fixed IMF. Conversely, the total mass-to-light ratio clearly 
increases with total stellar mass, possibly as a result of increased
dark matter or varying stellar IMF. The stellar M/L relations are all consistent with no intrinsic scatter, indicating a remarkable homogeneity in the stellar populations of these galaxies. The total M/L trends, on the other hand, exhibit 0.07-0.09 dex of intrinsic scatter; this is expected since the various parameter planes (\S~\ref{sec:3and4d}) demonstrate that three parameters are required to adequately describe ETGs \citep[e.g.,][]{graves}.

\begin{deluxetable*}{llllllll}
\tabletypesize{\scriptsize}
\tablecolumns{8}
\tablewidth{0pc}
\tablecaption{log[ M/L ] Linear Relations for SLACS Lenses}
\tablehead{
 & \multicolumn{3}{c}{Stellar M/L} && \multicolumn{3}{c}{Total M/L} \\
 \cline{2-4} \cline{6-8} \colhead{$X$} & \colhead{a} & \colhead{b} & \colhead{$\sigma_{\rm int}$} &&
 \colhead{a} & \colhead{b} & \colhead{$\sigma_{\rm int}$}  }
\startdata
\input{T_ml_fits.tex}
\enddata
\label{T_ml_fits}
\tablecomments{Fits are of the form log[ M/L / (M/L)$_\odot$] =
a*log$[ X ]$ + b and the stellar M/L is determined assuming a Chabrier
IMF. L$_B$ and L$_V$ are in units of $10^{10}$ L$_\odot$,
$\sigma_{e/2}$ is in units of 100 km s$^{-1}$, and the masses (M$_*$,
M$_{\rm dim}$, and \mre) are in units of $10^{11}$ M$_\odot$.}
\end{deluxetable*}

\begin{figure*}[ht]
\begin{center}
 \includegraphics[width=0.48\textwidth,clip]{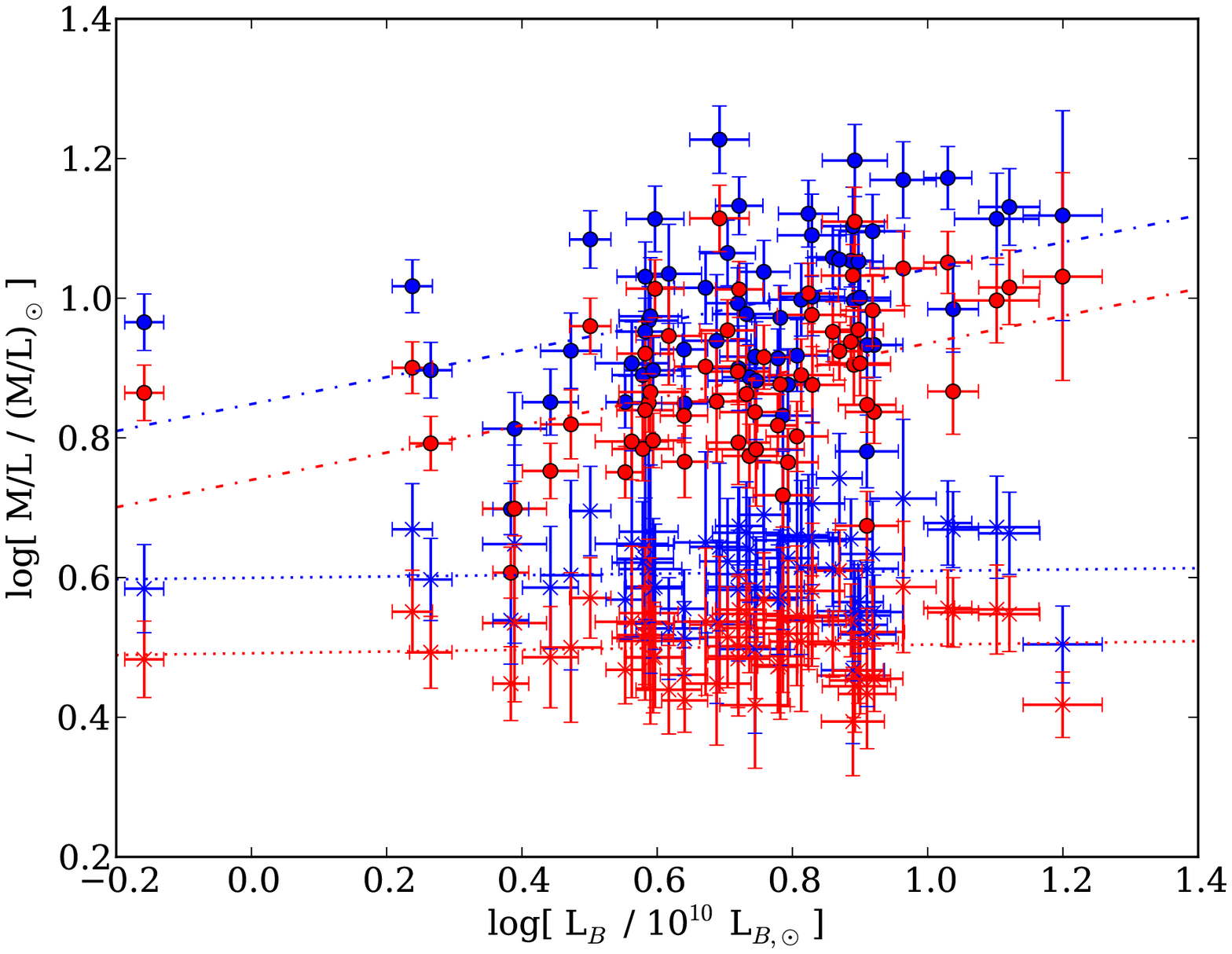}
 \includegraphics[width=0.48\textwidth,clip]{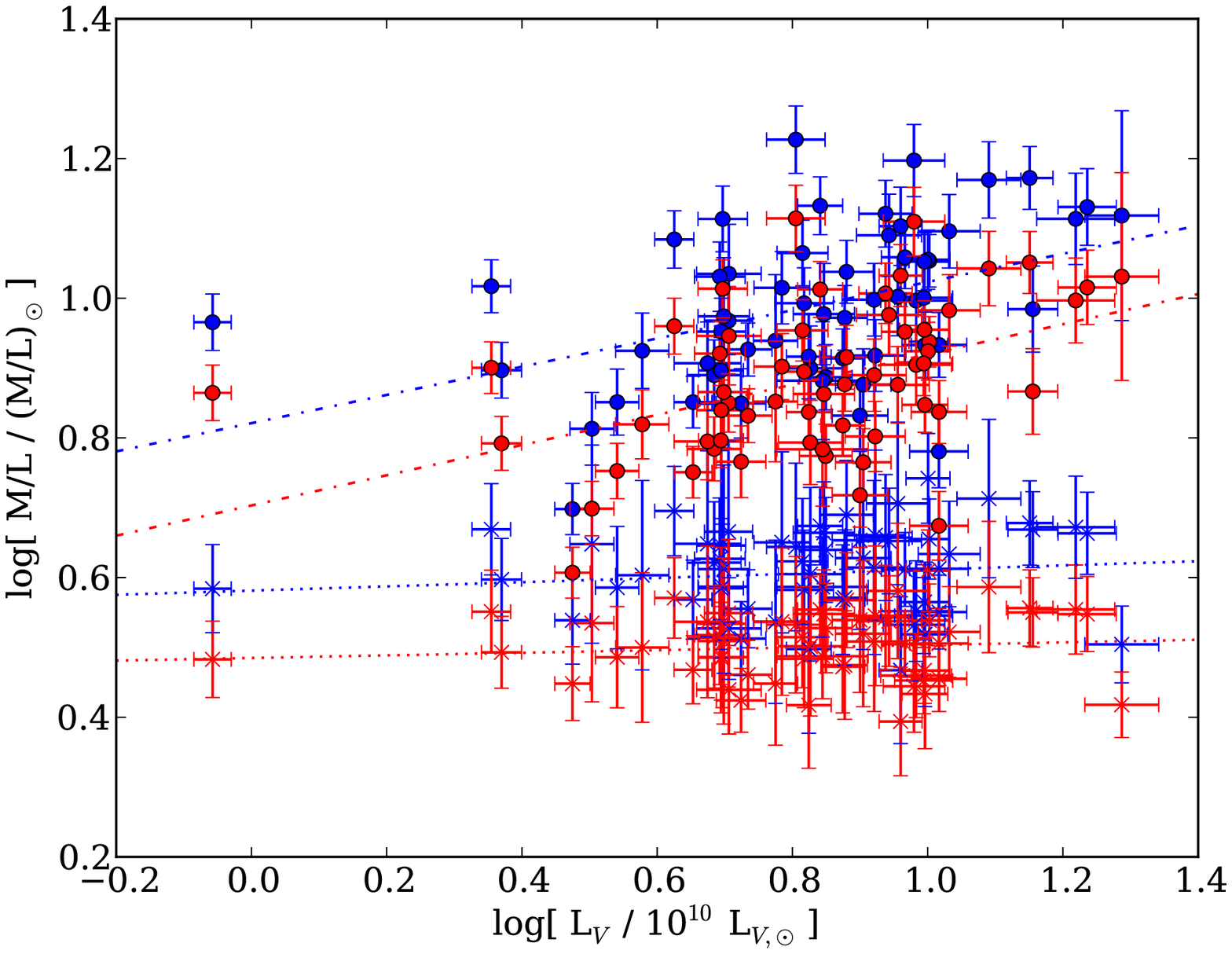}
 \includegraphics[width=0.48\textwidth,clip]{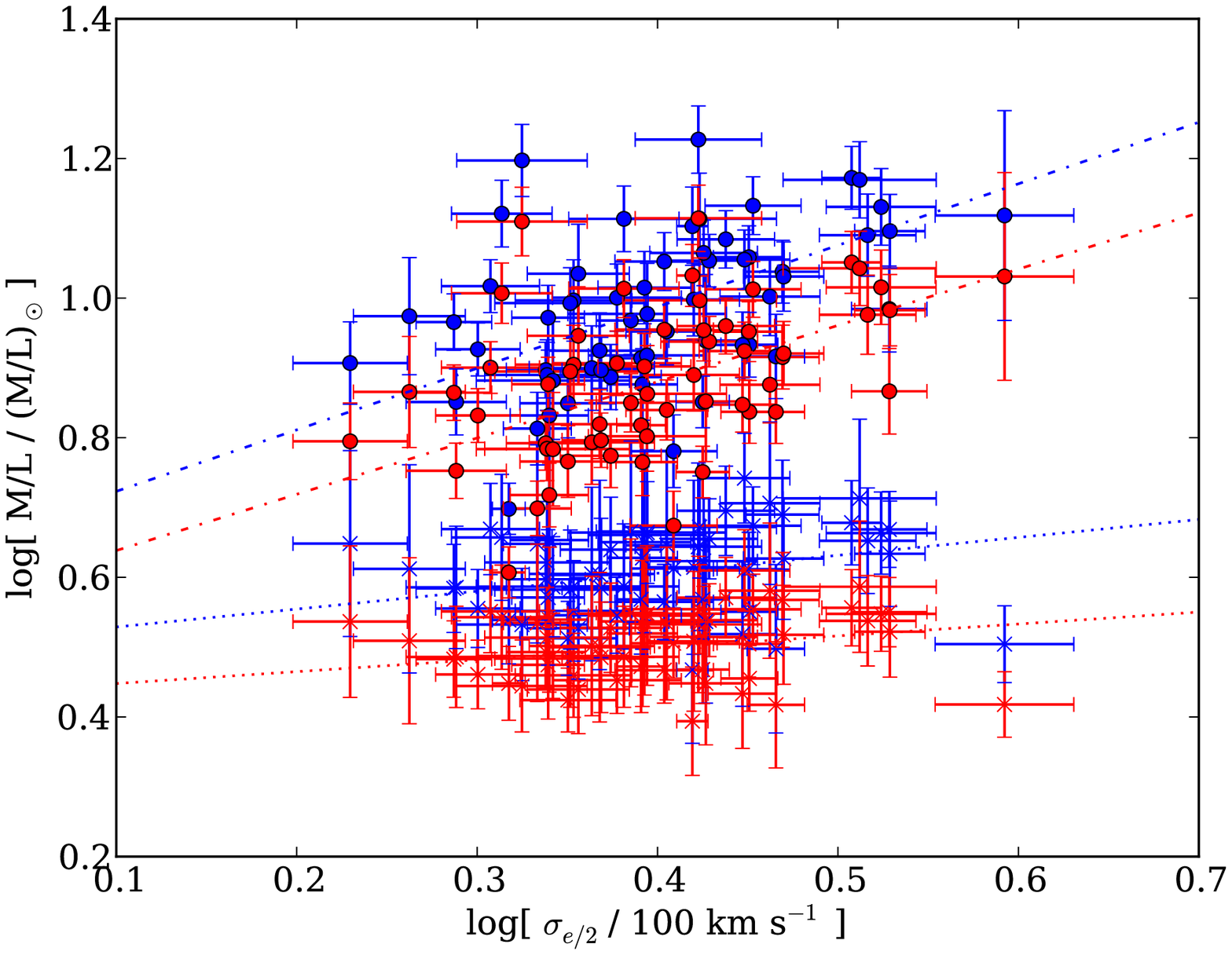}
 \includegraphics[width=0.48\textwidth,clip]{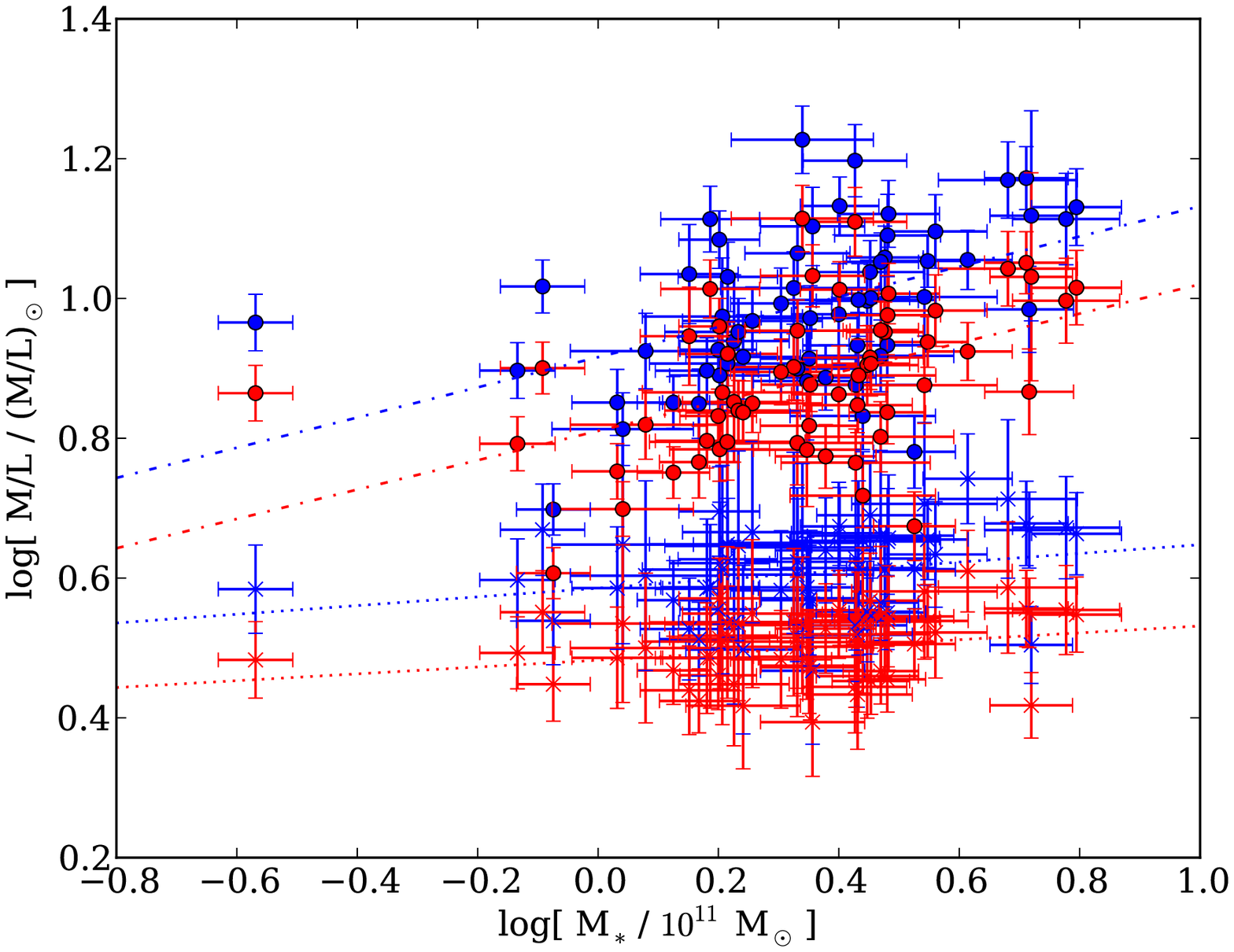}
 \includegraphics[width=0.48\textwidth,clip]{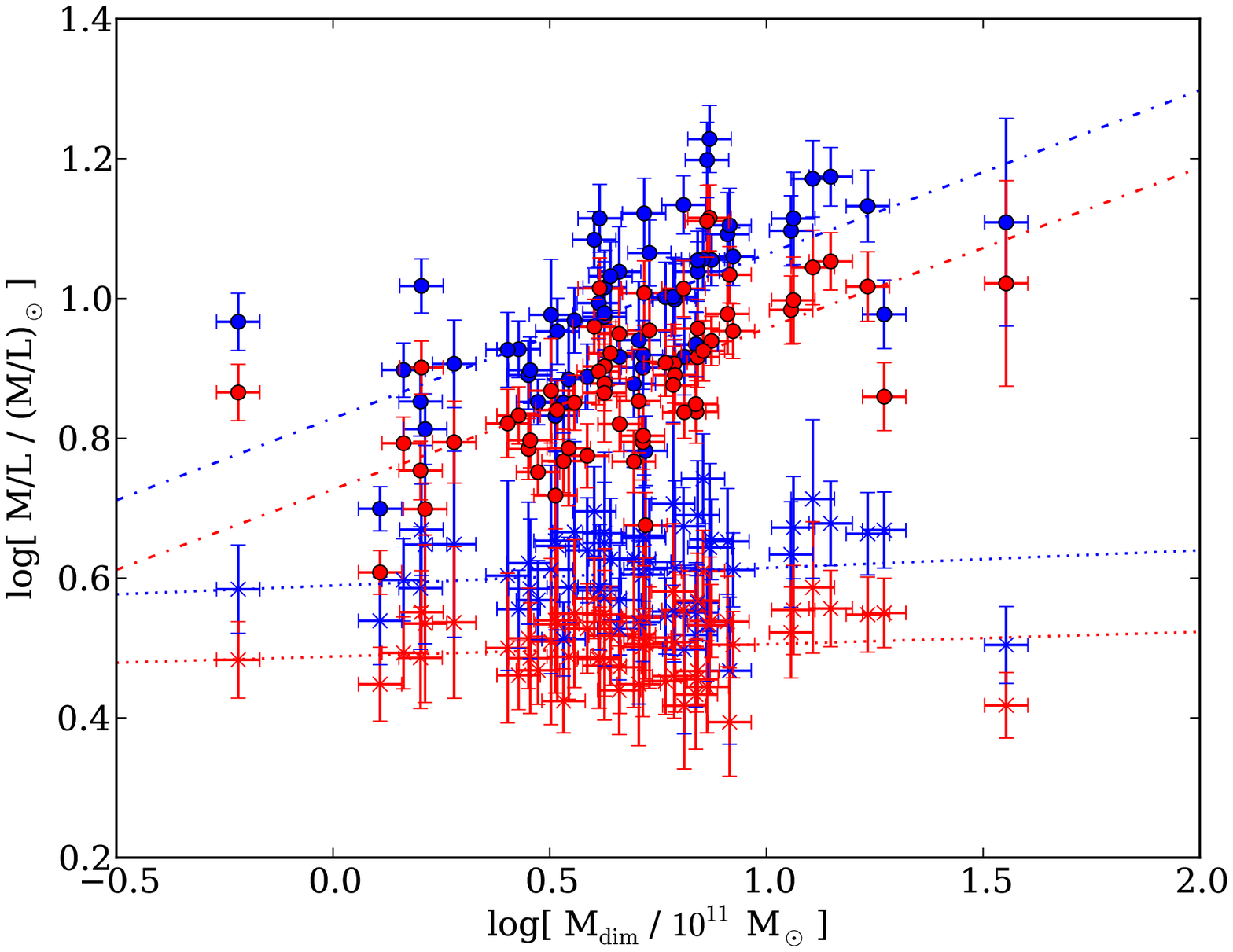}
 \includegraphics[width=0.48\textwidth,clip]{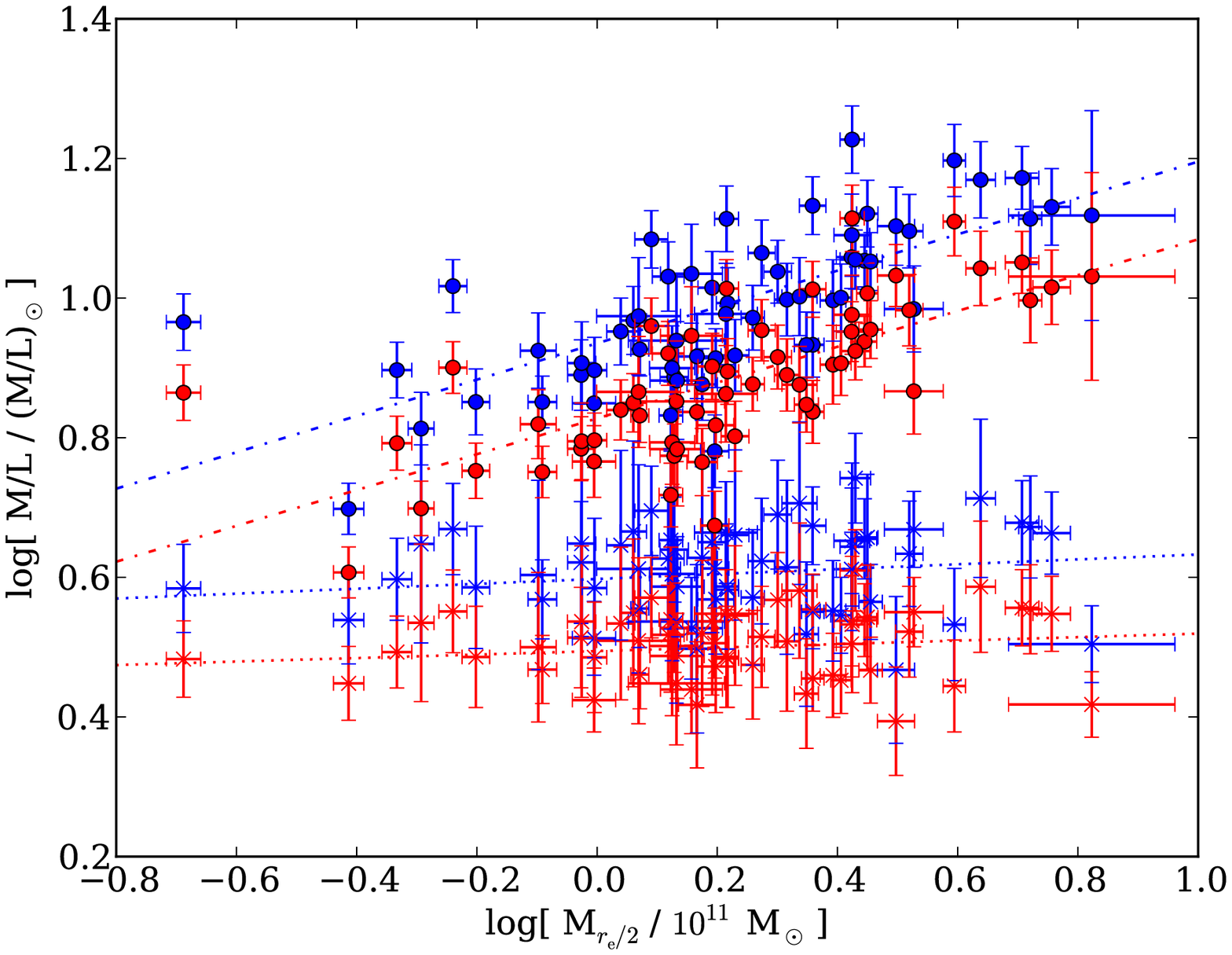}
\end{center}
 \caption{Relations between the stellar (crosses) and total (points)
 M/L in the $B$- (blue) and $V$- (red) bands for the SLACS lenses. A
 Chabrier IMF has been assumed for the stellar M/L. The dotted lines
 are linear fits to the stellar M/L relations and dash-dotted lines
 are linear fits to the total M/L relations.}
 \label{F_ml_fits}
\end{figure*}

In Figure~\ref{F_dm_fraction} we show the trends in the projected dark
matter fraction with the total mass inferred from lensing, the stellar
mass, the observed stellar velocity dispersion from SDSS, the
effective radius, and the $B$- and $V$-band luminosities evolved to $z
= 0$. Linear fits to these relations are provided in
Table~\ref{T_dm_fraction}. The most significant trends are with
effective radius and total mass \citep[also see][]{napolitano}, indicating that these parameters
govern the central dark matter fraction. These parameters are also consistent with having no intrinsic scatter at the 95\% confidence level (i.e., $\sigma_{\rm int}$ is within 2-$\sigma$ of 0), although the errors are large. We note that part of this scatter may also be due to the inadequacy of a linear fit; this is most clear for the Salpeter relations, which would indicate a physically impossible negative dark matter fraction if the linear trend is extrapolated below a stellar mass of $\sim 10^{10.5}$~M$_\odot$. A more appropriate model will be examined in a future paper (Auger et al.\ 2010, in preparation).

\begin{figure*}[ht]
\begin{center}
 \includegraphics[width=0.48\textwidth,clip]{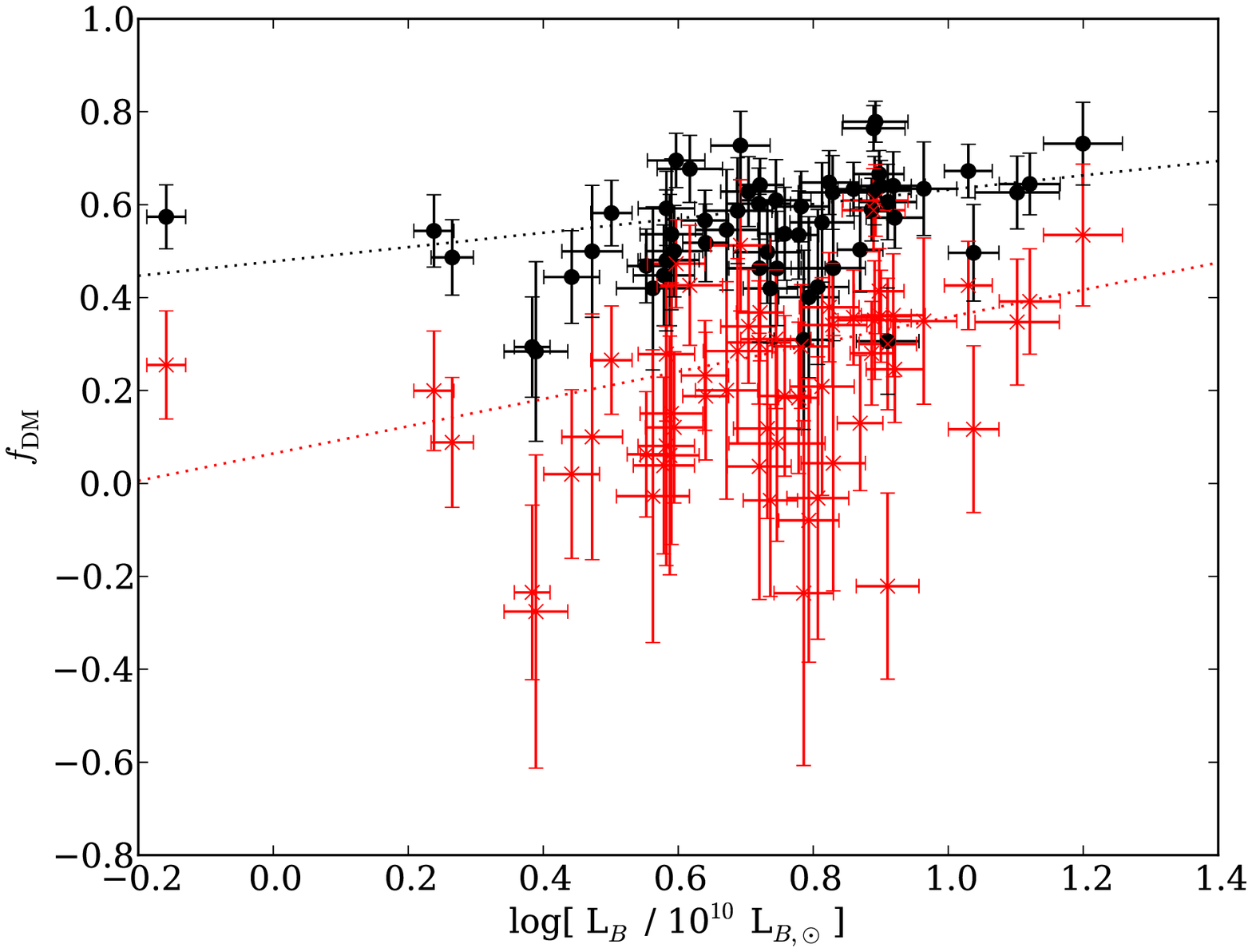}
 \includegraphics[width=0.48\textwidth,clip]{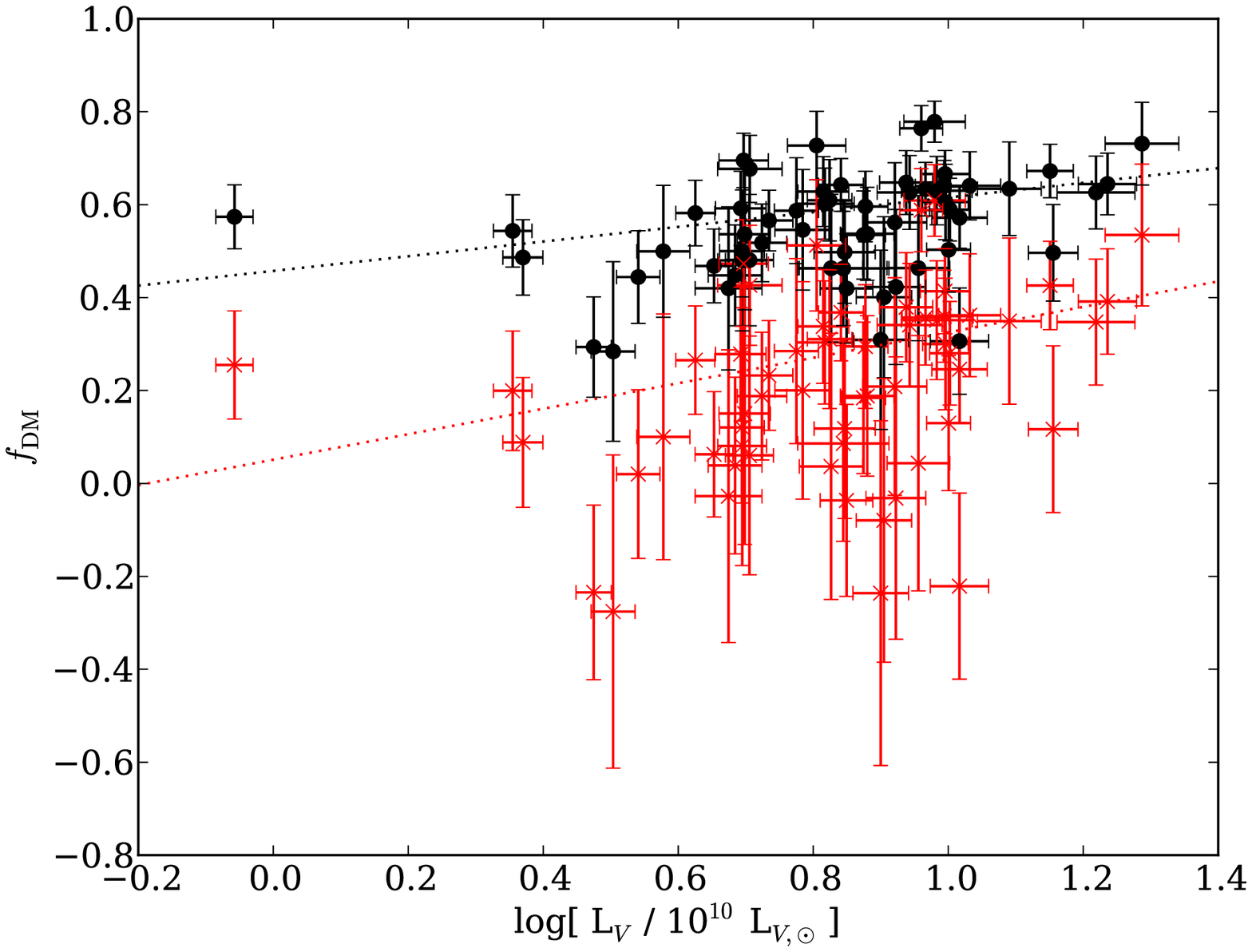}
 \includegraphics[width=0.48\textwidth,clip]{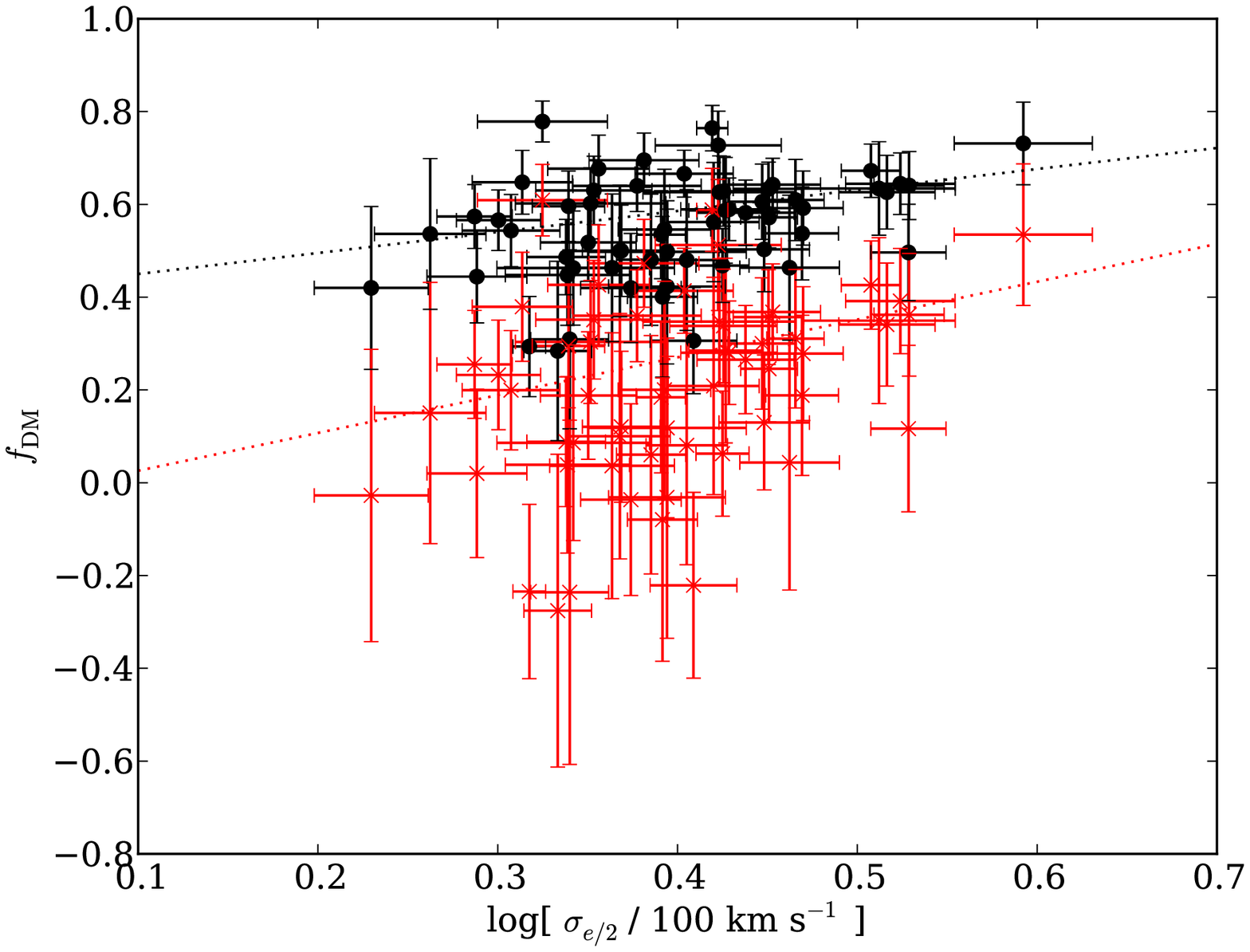}
 \includegraphics[width=0.48\textwidth,clip]{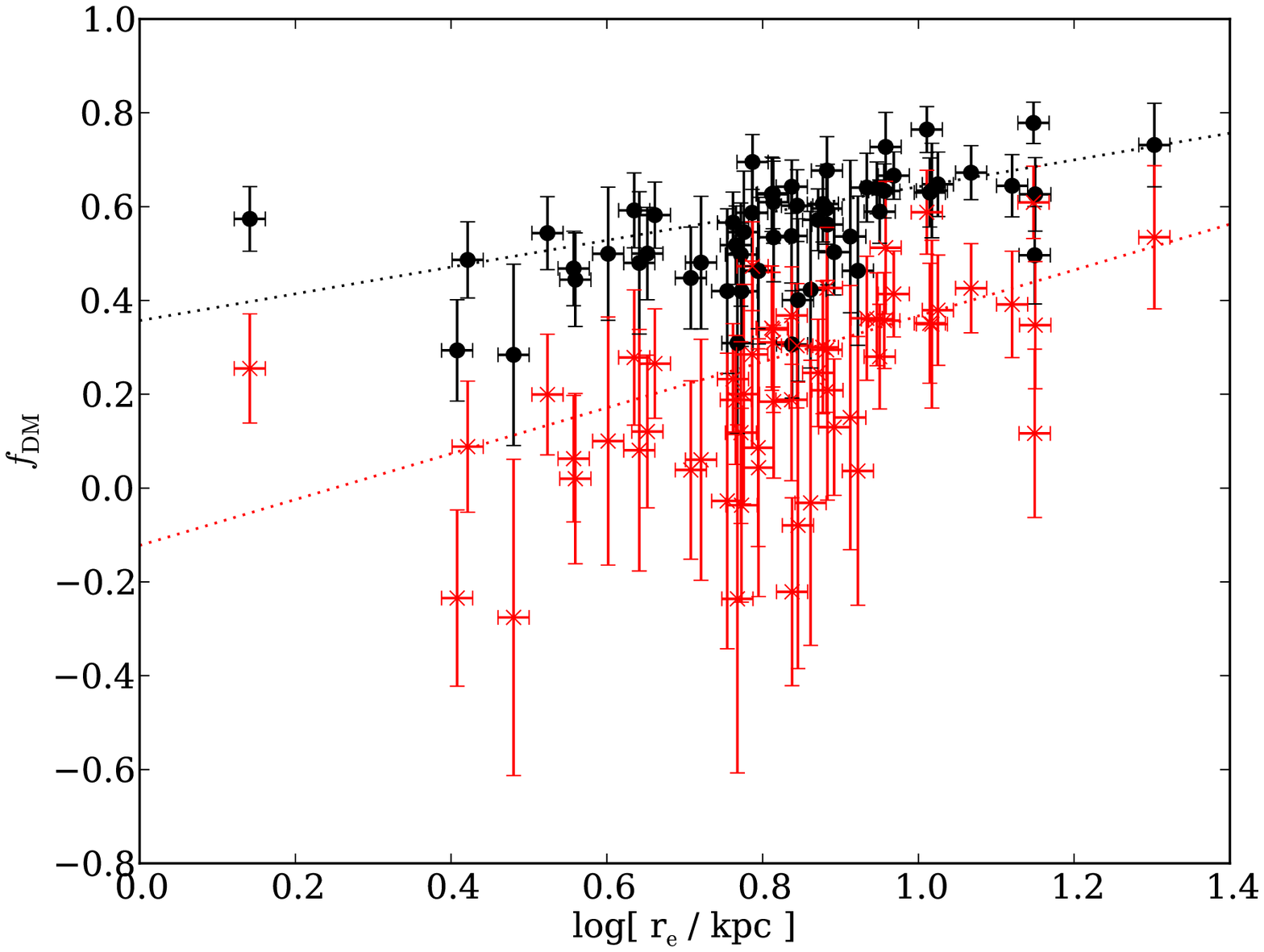}
 \includegraphics[width=0.48\textwidth,clip]{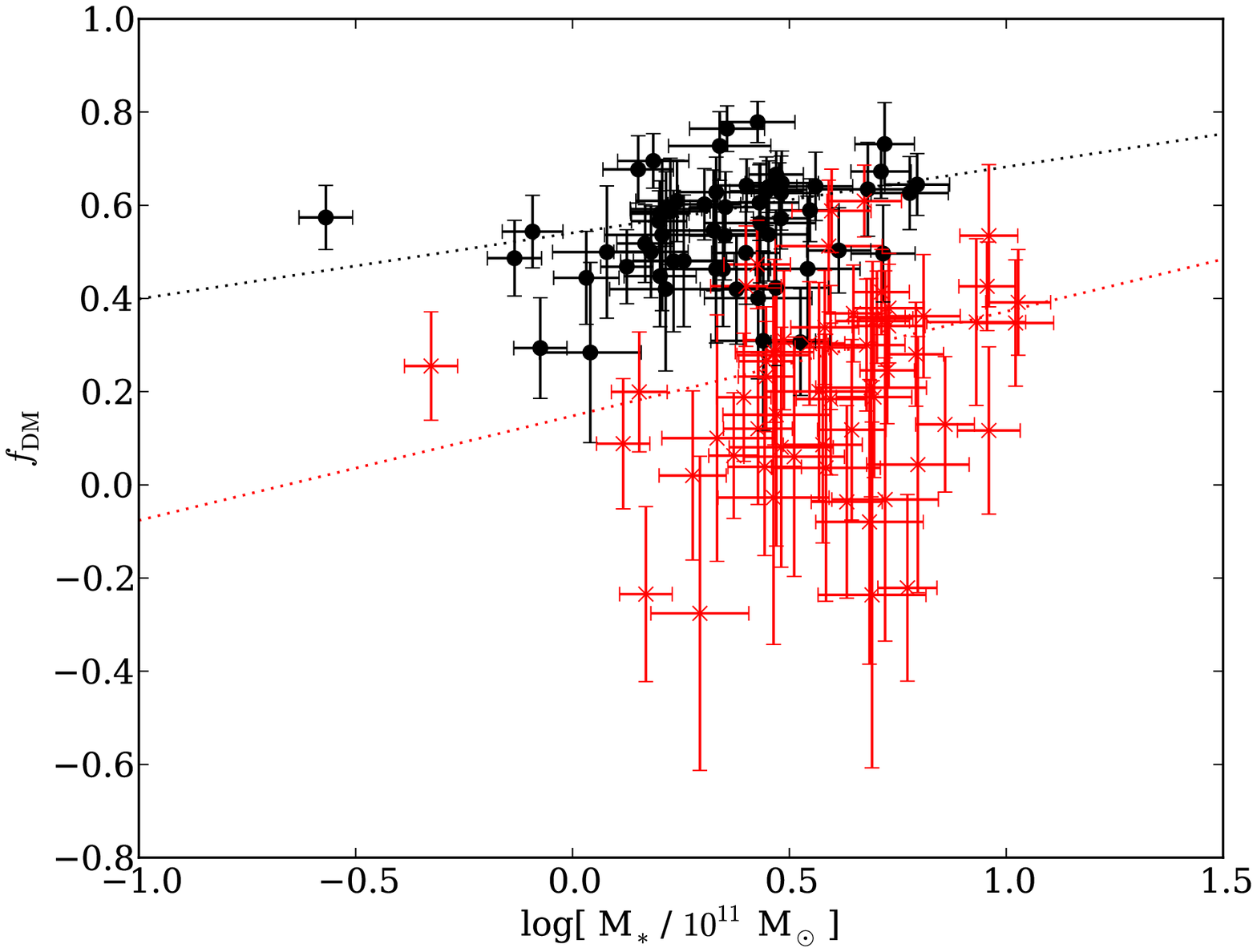}
 \includegraphics[width=0.48\textwidth,clip]{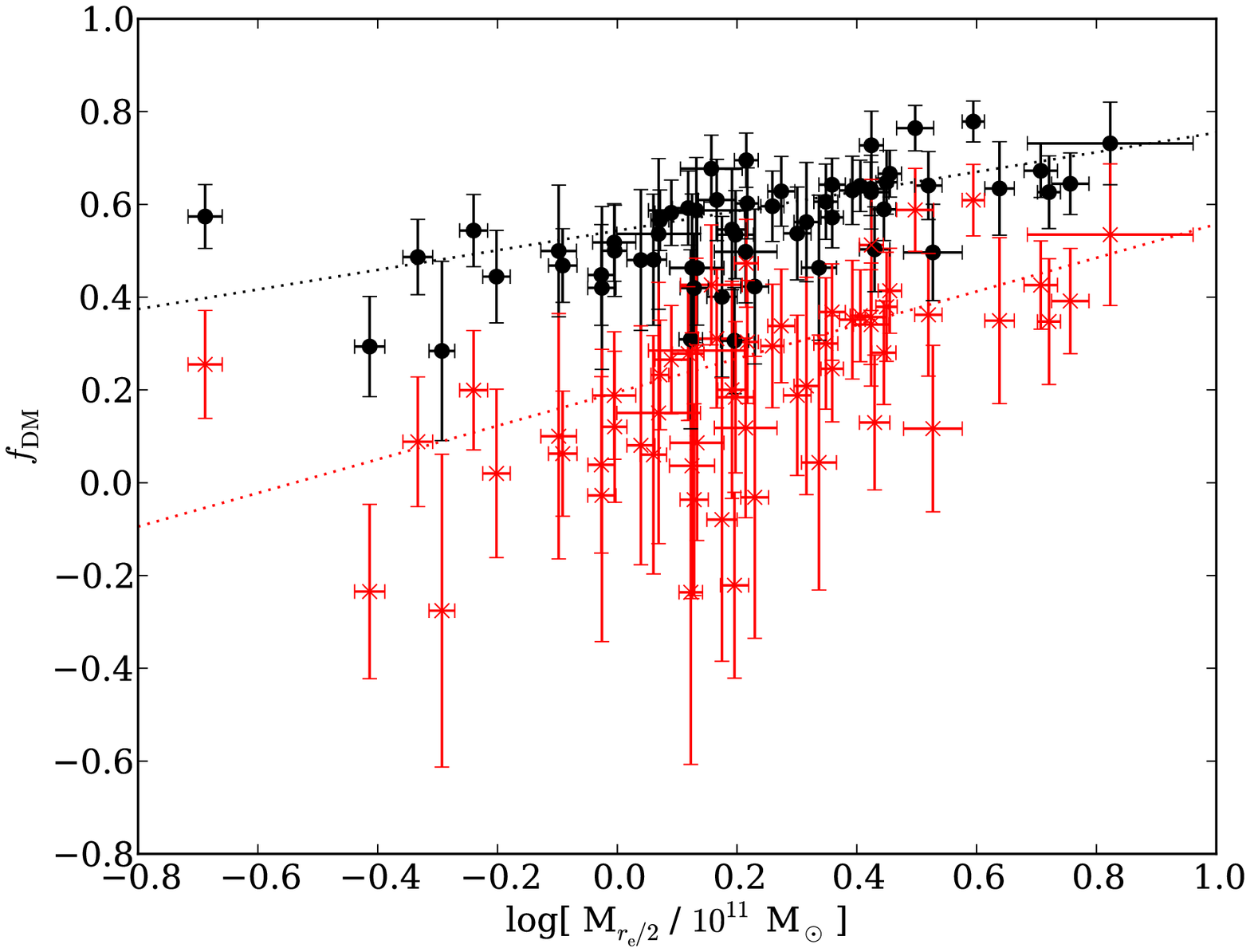}
\end{center}
 \caption{Relations between the projected dark matter fraction within
 half of the effective radius and L$_B$, L$_V$, $\sigma_{e/2}$,
 $r_{\rm e}$, M$_*$, and \mre. Red points are for a Salpeter IMF and
 black points are for a Chabrier IMF; the most
 significant trends are with $r_{\rm e}$ and \mre.}
 \label{F_dm_fraction}
\end{figure*}

\begin{deluxetable*}{llllllll}
\tabletypesize{\scriptsize}
\tablecolumns{8}
\tablewidth{0pc}
\tablecaption{$f_{\rm DM}$ Linear Relations for SLACS Lenses}
\tablehead{
 &
 \multicolumn{3}{c}{Chabrier IMF} &&
 \multicolumn{3}{c}{Salpeter IMF} \\ 
 \cline{2-4} \cline{6-8}
 \colhead{$X$} &
 \colhead{a} &
 \colhead{b} &
 \colhead{$\sigma_{\rm int}$} &&
 \colhead{a} &
 \colhead{b} &
 \colhead{$\sigma_{\rm int}$} 
}
\startdata
\input{T_dm_fits_2d.tex}
\enddata
\label{T_dm_fraction}
\tablecomments{Fits are of the form $f_{\rm DM}$ = a*log$[ X ]$ + b. L$_B$ and L$_V$ are in units of $10^{10}~L_\odot$, $\sigma_{e/2}$ is in units of 100~\kms, $r_{\rm e}$ is in kpc, and M$_*$ and \mre\ are in units of $10^{11}$~M$_\odot$.}
\end{deluxetable*}

\section{Correlations in three or more dimensions}
\label{sec:3and4d}

The SLACS lens galaxies have previously been shown to lie on the
Fundamental Plane \citep{slacsii,slacsvii} and on the Mass Plane
\citep{b07,slacsvii}. We now revisit these
relations with the enlarged SLACS sample and self-consistently evolved
luminosities, superceding our previous analysis. We also investigate
other scaling relations involving stellar mass, such as the stellar
mass including the Stellar Mass Plane (M$_*$P) and a new correlation
in a higher dimensional parameter space, which we call the Fundamental
Hyper-Plane.

\subsection{Fundamental and Mass Planes}
\label{ssec:3d}

We fit the parameter plane relations with the form
\be
\label{E_parameter_planes}
{\rm log}\ r_{\rm e}= \alpha^{\rm pp} {\rm log}\ \sigma_{e/2} + \beta^{\rm pp} {\rm log}\ \Lambda + \gamma^{\rm pp}, 
\ee 
where $\Lambda$ represents the average surface brightness within
$r_{\rm e}$, the average stellar mass surface density within $r_{\rm
e}$, or the average total mass surface density within $r_{\rm e}$/2.
The units used for the fits are: kpc for $r_{\rm e}$, 100~\kms for
$\sigma_{e/2}$, $10^9~L_\odot$ for the luminosity, $10^{9}~{\rm
M}_\odot$ for the stellar mass, and $10^{10}~{\rm M}_\odot$ for the
total mass. The intrinsic scatter is given in units of $\log r_{\rm
e}$. The inferred parameter planes are shown in Figure
\ref{F_parameter_planes} and illustrate the small intrinsic scatter
found in these relations.

We find that the FP relation is somewhat tilted with respect to
previous analyses \citep[e.g.,][]{slacsvii} if we allow the intrinsic
scatter to be a free parameter of our fit (see
Table~\ref{T_plane_parameters}). If we do not fit for intrinsic
scatter (that is, if we impose that the intrinsic scatter is zero) we
recover a FP consistent with other determinations
\citep{slacsvii,hydeMP}. This illustrates the importance of 
explicitly accounting for intrinsic scatter in the fit, even when it
is small like in the case of the FP. As a further consistency check we
repeated the fit of the FP parameters applying the $\sigma^{-4}$
scaling to account for the velocity dispersion selection of the SLACS
sample, and we find that the changes are insignificant. This is
consistent with the findings of \citet{hydeMP}, since our fitting
method is closer to their ``direct'' fit than to their orthogonal fit.
The non-zero scatter of the FP confirms previous results and is
consistent with being due to a combination of stellar population
differences and structural differences. The availability of stellar
mass and total mass diagnostics allows us to break this degeneracy as
we discuss in the rest of this section.

The coefficients for the M$_*$P plane are independent of the choice of
IMF for the SPS models (the normalization term changes by
$\approx0.25$, as is expected for Chabrier and Salpeter IMFs when
$\beta \approx -1$). It has previously been found that the M$_*$P
(where M$_*$ is inferred from SPS models) lies closer to the virial
plane ($\alpha=2$; $\beta=-1$) than the FP \citep{hydeMP}, although we
find in our data that the FP and M$_*$P are approximately aligned. We note,
however, that the M$_*$P is consistent with having no intrinsic
scatter ($\sigma_{\rm int} = 0.020\pm0.014$) while the FP has
intrinsic scatter of $\sigma_{\rm int} = 0.049\pm0.009$ dex; this
implies that the scatter in the FP is largely driven by small
differences in stellar populations (e.g., age or metallicity) for
massive galaxies, rather than from structural properties
(i.e., differences in dark matter content or the virial coefficient at a
fixed size and velocity dispersion).

The MP is also consistent with having no scatter (Table \ref{T_plane_parameters}) but is substantially mis-aligned with the FP and M$_*$P; instead, the MP is
approximately aligned with the virial plane 
(as was found in \citealt{b07,slacsvii}; also see \citealt{k09} where this plane included $\gamma^{\prime}$).
The offset with respect to the FP and M$_*$P is consistent with the bivariate relations (e.g., the middle panel of Figure \ref{F_mass_comparison} and the total M/L trends of Figure \ref{F_ml_fits}), while the slight offset from the virial plane may be related to the systematic deviation of the central mass density profile from isothermality (e.g., Figure \ref{F_gammap_correlations}) or anisotropy. The MP is also found to be consistent with no intrinsic scatter, although the errors are large.

\begin{deluxetable*}{lcccc}
\tabletypesize{\scriptsize}
\tablecolumns{5}
\tablewidth{0pc}
\tablecaption{Parameter Planes for SLACS Lenses}
\tablehead{
 \colhead{Plane} &
 \colhead{$\alpha^{\rm pp}$} &
 \colhead{$\beta^{\rm pp}$} &
 \colhead{$\gamma^{\rm pp}$} &
 \colhead{$\sigma_{\rm int}$}
}
\startdata
FP     &  $ 1.189\pm0.141$  &  $-0.885\pm0.041$  &   $-0.185\pm0.047$  &  \nodata  \\
M$_*$P &  $ 1.191\pm0.221$  &  $-0.971\pm0.073$  &   $\phantom{-}0.257\pm0.088$  &  \nodata  \\
MP     &  $ 1.829\pm0.133$  &  $-1.301\pm0.061$  &   $-0.301\pm0.055$  &  \nodata  \\ \\
FP     &  $ 1.020\pm0.203$  &  $-0.872\pm0.052$  &   $-0.108\pm0.076$  &  $0.049\pm0.009$  \\
M$_*$P &  $ 1.185\pm0.214$  &  $-0.952\pm0.074$  &   $\phantom{-}0.261\pm0.085$  &  $0.020\pm0.014$  \\
MP     &  $ 1.857\pm0.136$  &  $-1.279\pm0.065$  &   $-0.312\pm0.056$  &  $0.013\pm0.010$  \\

\enddata
\label{T_plane_parameters}
\tablecomments{Fits are of the form given in Equation \ref{E_parameter_planes}, with \reff\ in units of kpc, $\sigma_{e/2}$ in units of 100~km~s$^{-1}$, $V$-band luminosity and stellar mass in units of $10^9 {\rm L_\odot}$ and $10^9 {\rm M_\odot}$ respectively, and total mass in units of $10^{10} {\rm M_\odot}$. The first three fits are without intrinsic scatter while the latter three explicitly include scatter using the model of \citet{kelly}.}
\end{deluxetable*}
\vspace{1cm}

\begin{figure*}[ht]
\begin{center}
 \includegraphics[width=0.98\textwidth,clip]{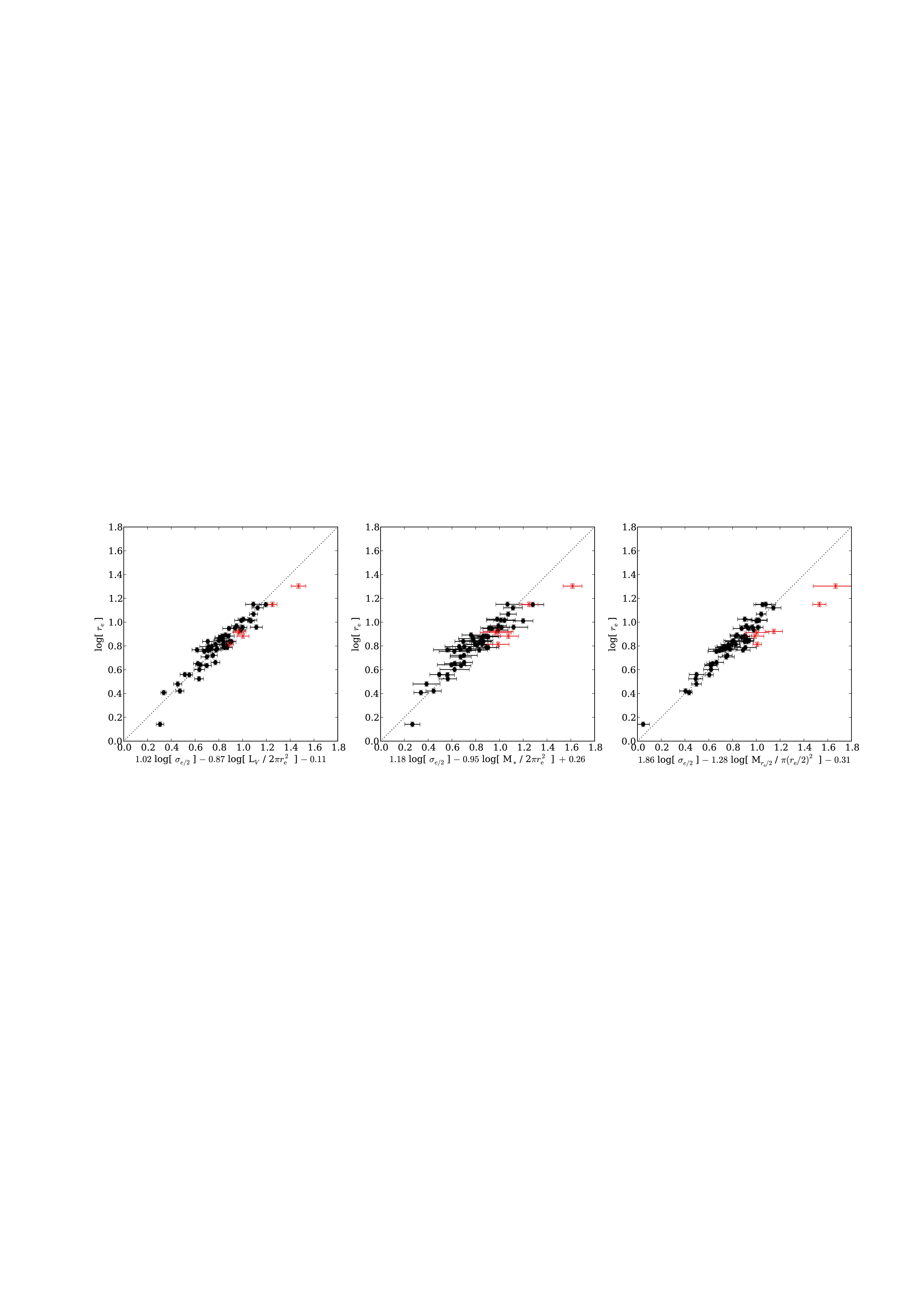}
\end{center}
\caption{FP, M$_*$P, and MP relations for the SLACS lenses. All three
planes show little intrinsic scatter ($\sigma_{\rm int} \lesssim
0.05$) and the FP and M$_*$P are well-aligned. The MP is found to be
offset from the FP and M$_*$P but is approximately aligned with the virial plane. The red points indicate the 6 galaxies which are outliers on the Fundamental Hyper-Plane and are not included in the fits.}
\label{F_parameter_planes}
\end{figure*}

\subsection{The Fundamental Hyper-Plane}
\label{S_manifold}

The M$_*$P and MP are found to have little intrinsic scatter but very
different orientations in size-velocity dispersion-mass space. We
would like to know whether the remaining scatter in these planes is
due to a non trivial relationship between \mre\ and M$_*$. We
therefore explicitly explore the relative importance of the stellar
and total mass on the structure of ETGs. In particular, we consider
the relationship between the effective radius, velocity dispersion,
central stellar mass, and central total mass by assuming a relation of
the form
\be
\label{E_hyperplane}
{\rm log}\ r_{\rm e} = \alpha^{\rm hp} {\rm log}\
\sigma_{e/2} + \beta^{\rm hp} {\rm log}\ \mmre + \gamma^{\rm hp} {\rm
log}\ {\rm M}_* + \delta^{\rm hp}
\ee
where M$_*$ is now the stellar mass within half the effective radius
to ensure consistency with the lensing determined total mass. An initial fit to this relation finds that a small number of galaxies, six in total, are significant outliers (greater than 3-$\sigma_{\rm int}$, where $\sigma_{\rm int} = 0.03$ for the initial fit). We re-fit the relation without these objects and find a substantially tighter relation with approximately the same coefficients but substantially decreased scatter. We therefore suspect that these objects have aberrant velocity dispersions and have therefore excluded them from our analysis, as discussed in Section 2. 

The best-fit relation for the Fundamental Hyper-Plane (FPH), with \reff measured in kpc, $\sigma_{e/2}$ in 100~km~s$^{-1}$, and both stellar and total mass in $10^{10} {\rm M_\odot}$, is given by
\bea
{\rm log}\ r_{\rm e} & = & (-0.91\pm0.10) {\rm log}\ \sigma_{e/2} + (0.69\pm0.04) {\rm log}\ \mmre + \nonumber \\
 &  &  (0.11\pm0.06) {\rm log}\ {\rm M}_* + (0.23\pm0.03)
\label{E_mass_structure}
\eea
with intrinsic scatter $\sigma_{\rm int} = 0.007\pm0.005$, and the fit is shown
in Figure \ref{F_manifold}. The dominant terms are the central velocity dispersion and central total mass, while the stellar mass only has a marginal role and is consistent with being unimportant at the 2-$\sigma$ level (that is, the coefficient for the M$_*$ term is within two-sigma of zero).

\begin{figure}
\begin{center}
 \includegraphics[width=0.48\textwidth,clip]{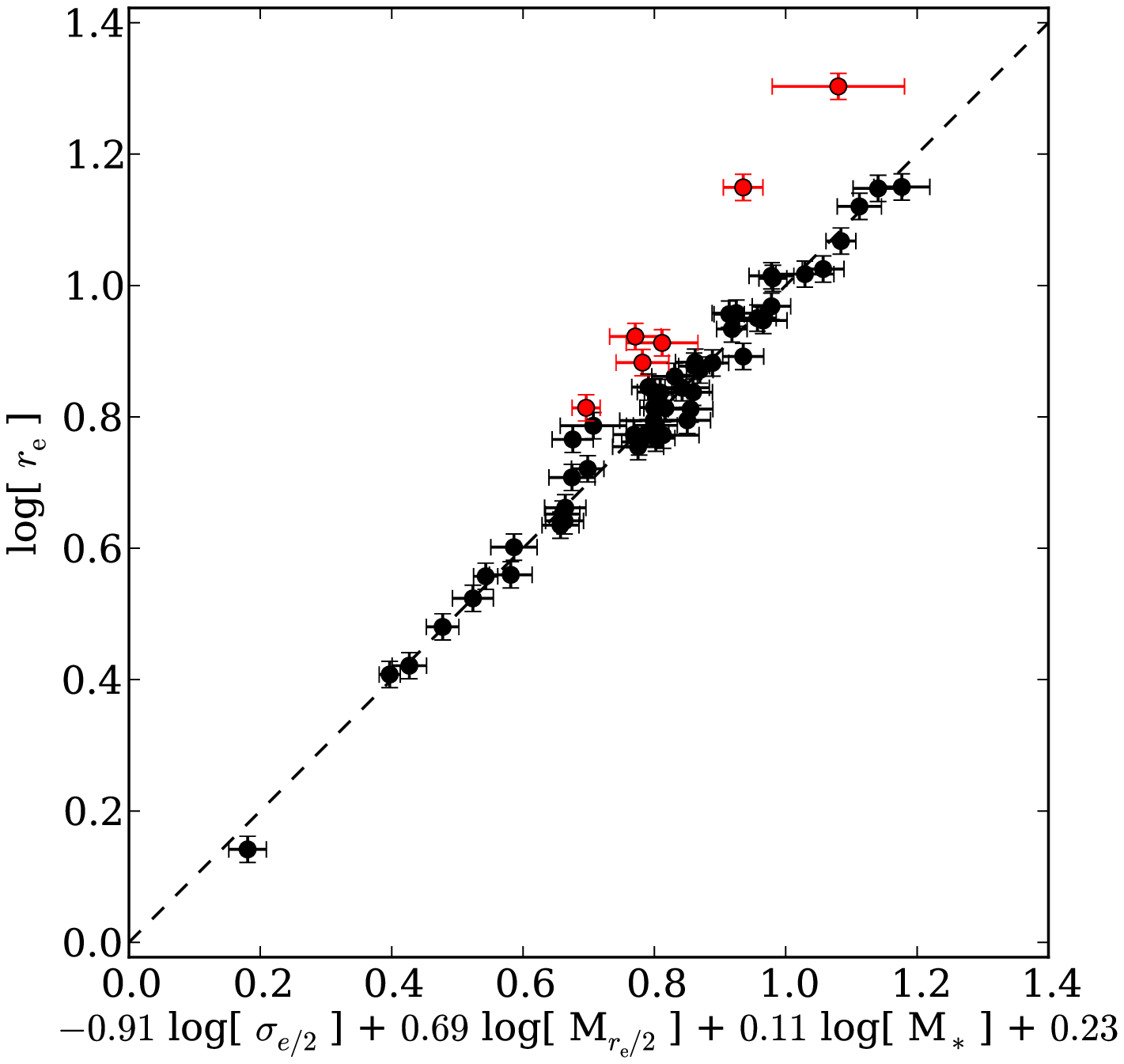}
\end{center}
 \caption{The relationship between the size, velocity dispersion, and mass (both stellar and total) of SLACS lenses. The red points are outliers that are rejected from the fit, which is found to be consistent with less than $3.5\%$ intrinsic scatter at 95\% confidence.}
 \label{F_manifold}
\end{figure}

\section{Discussion}

Before discussing our results in the broader context of studies of the
structure formation and evolution of ETGs, it is
essential to test whether the SLACS ETGs are indeed representative of
the overall population. In previous papers we have shown that SLACS lens
galaxies cannot be distinguished from control samples of ETGs selected
from the SDSS archive to have the same stellar velocity dispersion and
redshift \citep{slacsii,slacsv,slacsix}. In this paper we further
refine these tests by considering the complete SLACS sample of 73
ETGs, and testing a number of correlations, including the FP, the
M$_*$P and their projections, the M$_*$-\reff\ and M$_*$-$\sigma$
correlations. Once again we do not find any substantial difference
between the correlations inferred for our sample and for SDSS-selected
samples of non-lenses. Having found no significant evidence of a systematic
difference, we can safely assert that SLACS lenses are
representative of the entire population of massive ETGs and proceed to
interpret our results.

\subsection{The ``bulge-halo'' conspiracy, and implications of (small) departures from it}

The first key result of this study, building on previous SLACS
papers, is the precise measurement of the so-called ``bulge-halo
conspiracy'' and its tightness. In short, although mass clearly does
not follow light and an extended dark matter halo is needed to
reproduce simultaneously the lensing and dynamical constraints, the
two components add up to form almost exactly an isothermal total mass
density profile in the inner regions of galaxies ($\rho_{\rm tot}\propto r^{-\gamma'}$, with
$\gamma'=2$).  This is remarkable, since neither component is a single
power law, and there is no simple fundamental reason why this should
be the case, although general dynamical arguments based on incomplete
violent relaxation suggest that the isothermal sphere is a form of
dynamical attractor \citep[e.g.,][]{gunn,dekel,vanalbada,bertschinger,loeb}.
In addition, models based on simple prescriptions for baryonic
condensation \citep[e.g.,][]{gnedin} seem to suggest that it is possible to obtain
close-to-isothermal total mass density profiles starting from
cosmologically motivated dark matter halos \citep{jiang,humphrey}.

The resulting isothermal profile cannot be explained based purely on
dissipationless processes. In fact,
dissipationless processes in a cosmological setting would tend to produce inner density profiles
close to $\gamma'=1$ or even flatter \citep{navarro}.  However,
once the isothermal profile is established via dissipational processes
\citep[e.g.,][]{ciotti2007,robertson},
collisionless ``dry'' mergers preserve it quite accurately,
introducing a small amount of scatter, consistent with the observed
value \citep{NTB09}.  An interpretation of the
bulge-halo conspiracy and its possible origins are discussed at
lengths in previous SLACS papers \citep[e.g.,][]{slacsiii} and will
not be repeated here.

However, the precision achieved in this and previous SLACS studies \citep[e.g.,][]{slacsiii,k09}
allows us to highlight the importance of the
small but significant observed departures from the bulge-halo
conspiracy. Firstly, the average mass density profile is not exactly
isothermal but slightly steeper, $\gamma'=2.078\pm0.027$ \citep[this is marginally steeper than the slopes found by][who use X-ray temperature and density profiles to find the best-fit power-slope for the mass distribution of four ETGs]{humphrey}. Secondly,
there is evidence for non-negligible intrinsic scatter
$0.16\pm0.02$. Third, there is tentative evidence for a mild
dependency between $\gamma'$ on galaxy properties, such as radius or
central mass density.

The first fact has important implications for
gravitational lens studies, particularly those trying to infer
cosmography from gravitational time delays. Given the known degeneracy
between slope of the mass density profile and time delays
\citep[e.g.,][]{wucknitz}, if one assumes an isothermal prior then the
inferred Hubble Constant will typically be biased low by 10\% if no
other direct measure of the mass slope is available \citep[see][and references therein]{oguri,dobke,kochanekSF,TTreview}.
Additionally, due to the intrinsic scatter, estimates of the
Hubble Constant from a single lens can be off by 20\% if an isothermal
model is assumed and not independently constrained. Note, however, that these arguments are based on our assumption of isotropic orbits; anisotropy could change these biases considerably \citep[e.g.,][]{k09} and must be constrained in more detail in future work.
Nevertheless, additional external information on the mass density slope, such as that inferred from multiply-imaged extended sources \citep{warren,suyu06} and from stellar kinematics \citep{treu02,koopmans2003}, can help to break these degeneracies and provide robust estimates of cosmological parameters \citep[e.g.,][]{suyu}.

From the point of view of galaxy formation the intrinsic scatter and
the correlations between $\gamma'$ and galaxy properties are the most
interesting new elements \citep[also see][]{humphrey}. The
tightness imposes a constraint on the number of major merging events,
as well as on the diversity of formation histories. We will return to
this point later in \S~\ref{ssec:tight} in the context of the
tightness of the empirical correlation. The dependency of mass density
slope on galaxy parameters is another piece of evidence supporting the
idea that ETGs are not a homologous family from a structural point of
view. Both facts provide an interesting constraint for numerical
simulations that have sufficient resolution and baryonic physics to
simulate the inner regions of ETGs \citep[e.g.,][]{robertson,naab,gonzalezgarcia,lackner}. Small departures from regularity in the end may be the key to
understanding the details of ETGs formation \citep{kormendy09}.

\subsection{Non-triviality of empirical correlations}

The second key property of ETGs that we quantify in this study is the
degree to which their internal structure changes as a function of mass
or size. There are many ways to express this, including the observed
``tilt'' of the FP and the mass-dependent correlation between stellar mass
and dynamical mass. The lensing observables from the SLACS
sample allow us to add additional information and breaking some of the
degeneracies in the interpretation of these trends. We find that, over the range of
masses probed, the \emph{total} amount of mass is a non-linear
function of stellar mass, while the virial coefficient is
approximately constant. Similarly, we find that the trends cannot be
due to changes in stellar mass-to-light ratio for a fixed IMF. In
fact, the M$_*$/L inferred from SPS models is approximately the same
for all SLACS ETGs \citep[see also][]{tortora,grillo}, consistent with
the homogeneity of old stellar populations present in massive ETGs
\citep[e.g.,][]{thomas}. The fact that the {\it total} M/L varies
strongly as a function of mass, velocity dispersion, and luminosity
means that more massive galaxies have higher total M/L than less
massive systems. This suggests that either the central dark matter
fraction or the IMF is a strong function of mass \citep[e.g.,][]{t10,tortora}. In other words, it is the central fraction of dark matter, \emph{either
baryonic or non-baryonic}, that increases with galactic stellar mass.

These trends have long been connected with the presence of some
characteristic scale in the formation process of ETGs,
that may be due to baryonic physics. For example, the increase in cooling
timescales for the hot gas going from the lower mass to the higher
mass ETGs significantly changes the efficiency of converting baryons
into stars and therefore affects the dark matter fraction in the central
regions \citep[e.g.,][and references therein]{robertson}.
Additionally, changes in the modes of star formation that are responsible
for variations in chemical abundances could perhaps induce
variations in the stellar IMF \citep[but see][for a discussion]{graves}.
However, collisionless processes can also contribute to
the emergence of changes in structural properties with mass. For
example, the fraction of dark matter within the cylinder of radius
equal to a fixed fraction of the effective radius changes during dry
mergers as a result of the increase in effective radius \citep{NTB09}.

Interestingly, the observed trends cannot be explained by a
single phenomenon (for example, an increase in star formation efficiency with
total mass) as we know that ETGs occupy at least a two dimensional
subset of parameter space even when stellar population effects are
excluded by purely structural correlations like the Mass Plane and (assuming robust SPS models and ignoring the unknown normalization due to the IMF) the
Stellar Mass Plane. It appears that velocity dispersion as well as
size (or dynamical mass) are needed to fully specify the dynamical
properties of an ETG.  Furthermore, neither the MP nor
the M$_*$P appear to have intrinsic scatter. Thus, remarkably, it
appears that two parameters are not only necessary but also sufficient
to fully specify the internal structure of a massive ETG within our
observational errors. Finally, when we construct a direct relationship
between the structural parameters (effective radius, velocity
dispersion, stellar mass, and central total mass) of the SLACS lenses
(Equation \ref{E_mass_structure}) we find that the relationship is
nearly independent of the stellar mass. In terms of observables for non-lens
galaxies, the driving parameters are size and stellar velocity
dispersion, not stellar mass.

\subsection{Tightness of empirical correlations}
\label{ssec:tight}

The third key feature of ETGs addressed in this study is the tightness
of the empirical correlations. Traditional non-lensing correlations,
such as the FP, have small but non-zero intrinsic scatter \citep[e.g.,][]{jfk96,hydeMP}. \citet{graves} have recently emphasized the presence of
intrinsic scatter in the M$_*$P relation at the level of 0.02-0.03
dex, discussing several interpretations in terms of diversity in
stellar IMF and dark matter content. This low level of intrinsic
scatter cannot be ruled out by our data, although it should be noted
that our sample is restricted to the most massive systems and scatter
may be mass-dependent. Our study of correlations involving
total mass, including the MP, adds an important piece of evidence because the MP
is independent of SPS stellar mass estimates and therefore can be
used to break the degeneracy between the IMF and dark matter.
Our study shows that the MP is consistent with having
{\it no intrinsic scatter}, within the measurement errors of a few
hundredths of a dex ($\sigma_{\rm int} = 0.013\pm0.010$). Future studies comparing
the M$_*$P with the MP to even higher degrees of precision will be
able to quantify the contribution of IMF variations to the intrinsic
scatter of the M$_*$P.

The tightness of scaling relations is especially remarkable in a
scenario where evolution is driven by major mergers \citep[e.g.,][]{vanderwel}.
For example, dry mergers tend to move galaxies within
the FP and MP correlations and preserve their tightness. However, dry
mergers do not, in general, retain the tightness of the bivariate
projections of the parameter planes \citep{nipoti03,boylankolchin,nipoti}.
The properties of the progenitors
must be finely tuned with the orbital parameter of the merger in order
to produce the tight observed scaling relations.  \citet{nipoti} used
these relations to show that only half of the mass in ETGs at $z = 0$
can result from dry merging and dry merging cannot cause super-massive
galaxies at high redshifts \citep[e.g.,][]{trujillo,vandokkum} to
evolve into present-day ETGs.

\section{Summary}
\label{sec:summary}

We briefly summarize the most significant conclusions from our
analysis of the early-type lenses from the SLACS survey.

\begin{itemize}

\item The SLACS sample obeys all the standard correlations 
found for non-lensing ETGs, consistent with the hypothesis that it is
representative of velocity dispersion selected ETGs.

\item Stellar kinematics and lensing data constrain the
the slope of the total mass density profile ($\rho_{\rm tot}\propto
r^{-\gamma'}$). The average slope is found to be close to, but slightly
steeper than, isothermal, with $\langle \gamma' \rangle = 2.078\pm0.027$
and an intrinsic scatter of $0.16\pm0.02$.

\item The total mass density slope $\gamma'$ correlates with effective
radius (\reff) and central mass density ($\Sigma_{\rm tot}$) in the
sense that denser galaxies have steeper profiles. The residual intrinsic scatter is reduced but still significant ($0.12\pm0.02$ for $\Sigma_{\rm tot}$).

\item Tight correlations are found between dimensional mass 
M$_{\rm dim}=\frac{5 \sigma^2 r_{\rm e}}{G}$, stellar (M$_*$) and
total mass (M$_{\rm tot}$). The relationship between total mass and
dimensional mass is found to be consistent with linear with very little
scatter, implying that the virial coefficient of ETGs
is constant over this mass range. The correlation between total
(dynamical) mass and stellar mass is non linear (M$_* \propto \mmre^{0.8}$), consistent with the
hypothesis that the central CDM content and/or the
normalization of the stellar IMF changes with mass.

\item Assuming a universal initial mass function (IMF) the stellar
mass-to-light ratio is nearly constant over the range in masses probed
by the SLACS ETGs. In contrast, the total mass-to-light ratio
correlates strongly with lens properties; the most significant
correlations are with the central velocity dispersion and central
total mass. As a result, the dark matter fraction within \reff/2 is a
monotonically increasing function of galaxy mass and size.  If the
universal IMF assumption is relaxed, the trend could be explained at
least in part by an increasing IMF normalization with galaxy mass.

\item The Mass Plane (MP), obtained by replacing surface brightness
with surface mass density in the FP, is found to be tighter and closer
to the virial relation than the FP and the M$_*$P, indicating that the
scatter of those relations is dominated by stellar populations
effects. 

\item We construct the Fundamental Hyper-Plane by adding stellar
masses to the MP and find that the stellar mass coefficient is
consistent with zero and there is effectively no residual intrinsic
scatter.

\end{itemize}

Our results demonstrate that the dynamical structure of massive ETGs
is not scale invariant and that it is fully specified by size, stellar
velocity dispersion, and total mass.  Although the basic trends can be
explained qualitatively in terms of varying star formation efficiency
as a function of halo mass and as the result of dry and wet mergers,
reproducing quantitatively the observed correlations and their
tightness may be a significant challenge for galaxy formation models.
A detailed modeling effort will be presented in a follow-up paper,
where weak-lensing data will be combined with the present data to
strengthen the connection between the central part of the galaxies and
the virial mass of the halos in which their a embedded (Auger et al.\ 2010, in
preparation).

\acknowledgements
TT acknowledges support from the NSF thorough CAREER award NSF-0642621, by the Sloan Foundation through a Sloan Research Fellowship and by the Packard Foundation through a Packard Fellowship. LK is supported through an NWO-VIDI program subsidy (project number 639.042.505). RG acknowledges support from the Centre National des Etudes Spatiales. The work of LAM was carried out at the Jet Propulsion Laboratory, California Institute of Technology, under a contract with NASA. Support for programs \#10494, \#10798, and \#11202 was provided by NASA through grants from the Space Telescope Science Institute. STScI is operated by the Association of Universities for Research in Astronomy, Inc., under NASA contract NAS5-26555. Support for programs 10494, 10798, and 11202 was provided by NASA through grants from the Space Telescope Science Institute, This work has made use of the SDSS database. Funding for the SDSS and SDSS-II has been provided by the Alfred P. Sloan Foundation, the Participating Institutions, the National Science Foundation, the U.S. Department of Energy, the National Aeronautics and Space Administration, the Japanese Monbukagakusho, the Max Planck Society, and the Higher Education Funding Council for England.

\end{document}

%% file: T_lens_data.tex
SDSSJ0029$-$0055$\dagger$  &  \phantom{0}8.36  &  $231 \pm 18$  &  $11.33 \pm 0.13$  &  $11.58 \pm 0.13$  &  $11.13 \pm 0.03$  &  $ 0.47 \pm  0.15$  &  $\phantom{-}0.04 \pm 0.28$  &  $2.38 \pm 0.23$  \\
SDSSJ0037$-$0942  &  \phantom{0}7.44  &  $282 \pm 10$  &  $11.48 \pm 0.06$  &  $11.73 \pm 0.06$  &  $11.36 \pm 0.02$  &  $ 0.57 \pm  0.07$  &  $\phantom{-}0.25 \pm 0.11$  &  $2.14 \pm 0.07$  \\
SDSSJ0044$+$0113  &  \phantom{0}6.12  &  $267 \pm 13$  &  $11.23 \pm 0.09$  &  $11.47 \pm 0.09$  &  $11.13 \pm 0.07$  &  $ 0.59 \pm  0.10$  &  $\phantom{-}0.29 \pm 0.18$  &  $2.31 \pm 0.24$  \\
SDSSJ0216$-$0813  &  13.19  &  $334 \pm 23$  &  $11.79 \pm 0.07$  &  $12.03 \pm 0.07$  &  $11.76 \pm 0.02$  &  $ 0.65 \pm  0.06$  &  $\phantom{-}0.40 \pm 0.11$  &  $2.09 \pm 0.20$  \\
SDSSJ0252$+$0039  &  \phantom{0}5.68  &  $170 \pm 12$  &  $11.21 \pm 0.13$  &  $11.46 \pm 0.13$  &  $10.97 \pm 0.03$  &  $ 0.42 \pm  0.18$  &  $-0.03 \pm 0.32$  &  $1.57 \pm 0.12$  \\
SDSSJ0330$-$0020  &  \phantom{0}6.23  &  $220 \pm 21$  &  $11.35 \pm 0.09$  &  $11.58 \pm 0.09$  &  $11.14 \pm 0.05$  &  $ 0.46 \pm  0.12$  &  $\phantom{-}0.09 \pm 0.21$  &  $1.91 \pm 0.18$  \\
SDSSJ0728$+$3835  &  \phantom{0}5.86  &  $219 \pm 11$  &  $11.44 \pm 0.12$  &  $11.69 \pm 0.12$  &  $11.12 \pm 0.02$  &  $ 0.31 \pm  0.19$  &  $-0.23 \pm 0.37$  &  $1.86 \pm 0.10$  \\
SDSSJ0737$+$3216$\dagger$  &  14.10  &  $338 \pm 16$  &  $11.72 \pm 0.07$  &  $11.96 \pm 0.07$  &  $11.52 \pm 0.03$  &  $ 0.49 \pm  0.09$  &  $\phantom{-}0.10 \pm 0.16$  &  $2.68 \pm 0.12$  \\
SDSSJ0819$+$4534$\dagger$  &  \phantom{0}7.63  &  $227 \pm 15$  &  $11.15 \pm 0.08$  &  $11.40 \pm 0.08$  &  $11.16 \pm 0.04$  &  $ 0.68 \pm  0.07$  &  $\phantom{-}0.43 \pm 0.12$  &  $2.16 \pm 0.25$  \\
SDSSJ0822$+$2652  &  \phantom{0}7.64  &  $263 \pm 15$  &  $11.43 \pm 0.13$  &  $11.69 \pm 0.13$  &  $11.32 \pm 0.02$  &  $ 0.56 \pm  0.13$  &  $\phantom{-}0.21 \pm 0.23$  &  $2.12 \pm 0.14$  \\
SDSSJ0912$+$0029  &  11.69  &  $322 \pm 12$  &  $11.71 \pm 0.07$  &  $11.96 \pm 0.07$  &  $11.71 \pm 0.02$  &  $ 0.67 \pm  0.06$  &  $\phantom{-}0.43 \pm 0.09$  &  $1.98 \pm 0.09$  \\
SDSSJ0935$-$0003$\dagger$  &  20.09  &  $391 \pm 35$  &  $11.72 \pm 0.07$  &  $11.96 \pm 0.07$  &  $11.81 \pm 0.14$  &  $ 0.73 \pm  0.09$  &  $\phantom{-}0.53 \pm 0.15$  &  $2.44 \pm 0.36$  \\
SDSSJ0936$+$0913  &  \phantom{0}7.00  &  $246 \pm 11$  &  $11.43 \pm 0.12$  &  $11.68 \pm 0.12$  &  $11.18 \pm 0.02$  &  $ 0.40 \pm  0.17$  &  $-0.07 \pm 0.30$  &  $2.24 \pm 0.12$  \\
SDSSJ0946$+$1006  &  \phantom{0}9.08  &  $265 \pm 21$  &  $11.34 \pm 0.12$  &  $11.59 \pm 0.12$  &  $11.43 \pm 0.02$  &  $ 0.73 \pm  0.07$  &  $\phantom{-}0.51 \pm 0.14$  &  $2.01 \pm 0.18$  \\
SDSSJ0956$+$5100  &  \phantom{0}8.58  &  $338 \pm 15$  &  $11.56 \pm 0.09$  &  $11.81 \pm 0.08$  &  $11.52 \pm 0.02$  &  $ 0.64 \pm  0.07$  &  $\phantom{-}0.36 \pm 0.13$  &  $2.30 \pm 0.09$  \\
SDSSJ0959$+$0410  &  \phantom{0}3.34  &  $203 \pm 13$  &  $10.91 \pm 0.07$  &  $11.15 \pm 0.06$  &  $10.76 \pm 0.02$  &  $ 0.54 \pm  0.08$  &  $\phantom{-}0.20 \pm 0.13$  &  $2.05 \pm 0.15$  \\
SDSSJ0959$+$4416  &  \phantom{0}7.27  &  $248 \pm 19$  &  $11.47 \pm 0.12$  &  $11.72 \pm 0.12$  &  $11.23 \pm 0.02$  &  $ 0.42 \pm  0.17$  &  $-0.03 \pm 0.30$  &  $2.14 \pm 0.21$  \\
SDSSJ1016$+$3859  &  \phantom{0}4.38  &  $254 \pm 13$  &  $11.23 \pm 0.12$  &  $11.48 \pm 0.12$  &  $11.04 \pm 0.02$  &  $ 0.48 \pm  0.15$  &  $\phantom{-}0.08 \pm 0.26$  &  $2.19 \pm 0.11$  \\
SDSSJ1020$+$1122  &  \phantom{0}6.23  &  $290 \pm 18$  &  $11.54 \pm 0.12$  &  $11.80 \pm 0.12$  &  $11.34 \pm 0.03$  &  $ 0.46 \pm  0.16$  &  $\phantom{-}0.04 \pm 0.28$  &  $2.08 \pm 0.12$  \\
SDSSJ1023$+$4230  &  \phantom{0}5.97  &  $247 \pm 15$  &  $11.33 \pm 0.12$  &  $11.57 \pm 0.12$  &  $11.19 \pm 0.03$  &  $ 0.55 \pm  0.13$  &  $\phantom{-}0.20 \pm 0.23$  &  $2.01 \pm 0.11$  \\
SDSSJ1029$+$0420  &  \phantom{0}3.02  &  $215 \pm \phantom{0}9$  &  $11.04 \pm 0.12$  &  $11.29 \pm 0.11$  &  $10.71 \pm 0.02$  &  $ 0.28 \pm  0.19$  &  $-0.28 \pm 0.34$  &  $2.28 \pm 0.10$  \\
SDSSJ1106$+$5228  &  \phantom{0}3.61  &  $266 \pm \phantom{0}9$  &  $11.13 \pm 0.06$  &  $11.37 \pm 0.06$  &  $10.91 \pm 0.02$  &  $ 0.47 \pm  0.08$  &  $\phantom{-}0.07 \pm 0.13$  &  $2.40 \pm 0.07$  \\
SDSSJ1112$+$0826  &  \phantom{0}6.48  &  $328 \pm 20$  &  $11.48 \pm 0.09$  &  $11.73 \pm 0.08$  &  $11.43 \pm 0.03$  &  $ 0.63 \pm  0.08$  &  $\phantom{-}0.34 \pm 0.13$  &  $2.21 \pm 0.10$  \\
SDSSJ1134$+$6027  &  \phantom{0}5.26  &  $243 \pm 11$  &  $11.26 \pm 0.12$  &  $11.51 \pm 0.12$  &  $11.06 \pm 0.02$  &  $ 0.48 \pm  0.14$  &  $\phantom{-}0.06 \pm 0.25$  &  $2.20 \pm 0.11$  \\
SDSSJ1142$+$1001  &  \phantom{0}6.99  &  $225 \pm 22$  &  $11.30 \pm 0.08$  &  $11.55 \pm 0.08$  &  $11.22 \pm 0.02$  &  $ 0.60 \pm  0.08$  &  $\phantom{-}0.31 \pm 0.13$  &  $1.90 \pm 0.23$  \\
SDSSJ1143$-$0144  &  10.25  &  $263 \pm \phantom{0}5$  &  $11.36 \pm 0.09$  &  $11.60 \pm 0.09$  &  $11.50 \pm 0.03$  &  $ 0.77 \pm  0.05$  &  $\phantom{-}0.59 \pm 0.09$  &  $1.92 \pm 0.06$  \\
SDSSJ1153$+$4612  &  \phantom{0}4.00  &  $233 \pm 15$  &  $11.08 \pm 0.13$  &  $11.33 \pm 0.13$  &  $10.90 \pm 0.03$  &  $ 0.50 \pm  0.14$  &  $\phantom{-}0.10 \pm 0.26$  &  $2.28 \pm 0.13$  \\
SDSSJ1204$+$0358  &  \phantom{0}4.59  &  $274 \pm 17$  &  $11.20 \pm 0.07$  &  $11.45 \pm 0.06$  &  $11.09 \pm 0.03$  &  $ 0.58 \pm  0.07$  &  $\phantom{-}0.26 \pm 0.12$  &  $2.29 \pm 0.11$  \\
SDSSJ1205$+$4910  &  \phantom{0}9.04  &  $282 \pm 13$  &  $11.48 \pm 0.06$  &  $11.72 \pm 0.06$  &  $11.42 \pm 0.02$  &  $ 0.63 \pm  0.06$  &  $\phantom{-}0.36 \pm 0.10$  &  $2.16 \pm 0.12$  \\
SDSSJ1213$+$6708$\dagger$  &  \phantom{0}6.51  &  $292 \pm 11$  &  $11.24 \pm 0.10$  &  $11.49 \pm 0.09$  &  $11.17 \pm 0.02$  &  $ 0.61 \pm  0.08$  &  $\phantom{-}0.31 \pm 0.14$  &  $2.49 \pm 0.08$  \\
SDSSJ1218$+$0830  &  \phantom{0}7.62  &  $218 \pm 10$  &  $11.35 \pm 0.08$  &  $11.59 \pm 0.08$  &  $11.26 \pm 0.02$  &  $ 0.60 \pm  0.08$  &  $\phantom{-}0.30 \pm 0.13$  &  $1.82 \pm 0.11$  \\
SDSSJ1250$+$0523  &  \phantom{0}6.88  &  $256 \pm 14$  &  $11.53 \pm 0.07$  &  $11.77 \pm 0.07$  &  $11.20 \pm 0.02$  &  $ 0.31 \pm  0.11$  &  $-0.22 \pm 0.20$  &  $2.30 \pm 0.12$  \\
SDSSJ1306$+$0600  &  \phantom{0}6.12  &  $241 \pm 17$  &  $11.19 \pm 0.08$  &  $11.43 \pm 0.08$  &  $11.22 \pm 0.02$  &  $ 0.70 \pm  0.06$  &  $\phantom{-}0.47 \pm 0.10$  &  $1.89 \pm 0.14$  \\
SDSSJ1313$+$4615  &  \phantom{0}6.51  &  $266 \pm 18$  &  $11.33 \pm 0.09$  &  $11.58 \pm 0.08$  &  $11.27 \pm 0.02$  &  $ 0.63 \pm  0.08$  &  $\phantom{-}0.34 \pm 0.12$  &  $2.06 \pm 0.14$  \\
SDSSJ1318$-$0313  &  14.05  &  $211 \pm 18$  &  $11.43 \pm 0.09$  &  $11.67 \pm 0.09$  &  $11.60 \pm 0.02$  &  $ 0.78 \pm  0.04$  &  $\phantom{-}0.61 \pm 0.08$  &  $1.64 \pm 0.15$  \\
SDSSJ1330$-$0148  &  \phantom{0}1.39  &  $194 \pm \phantom{0}9$  &  $10.43 \pm 0.06$  &  $10.67 \pm 0.06$  &  $10.31 \pm 0.03$  &  $ 0.57 \pm  0.07$  &  $\phantom{-}0.26 \pm 0.12$  &  $2.25 \pm 0.10$  \\
SDSSJ1402$+$6321  &  \phantom{0}8.92  &  $268 \pm 17$  &  $11.55 \pm 0.07$  &  $11.79 \pm 0.06$  &  $11.45 \pm 0.02$  &  $ 0.59 \pm  0.07$  &  $\phantom{-}0.28 \pm 0.11$  &  $1.97 \pm 0.14$  \\
SDSSJ1403$+$0006  &  \phantom{0}5.10  &  $218 \pm 17$  &  $11.20 \pm 0.08$  &  $11.44 \pm 0.08$  &  $10.97 \pm 0.02$  &  $ 0.45 \pm  0.11$  &  $\phantom{-}0.04 \pm 0.19$  &  $2.14 \pm 0.23$  \\
SDSSJ1416$+$5136  &  \phantom{0}5.92  &  $248 \pm 25$  &  $11.40 \pm 0.08$  &  $11.64 \pm 0.08$  &  $11.22 \pm 0.05$  &  $ 0.50 \pm  0.11$  &  $\phantom{-}0.12 \pm 0.20$  &  $1.90 \pm 0.16$  \\
SDSSJ1420$+$6019  &  \phantom{0}2.56  &  $208 \pm \phantom{0}4$  &  $10.93 \pm 0.06$  &  $11.17 \pm 0.06$  &  $10.59 \pm 0.02$  &  $ 0.30 \pm  0.10$  &  $-0.23 \pm 0.18$  &  $2.28 \pm 0.07$  \\
SDSSJ1430$+$4105  &  10.41  &  $325 \pm 32$  &  $11.68 \pm 0.12$  &  $11.93 \pm 0.11$  &  $11.64 \pm 0.02$  &  $ 0.63 \pm  0.10$  &  $\phantom{-}0.35 \pm 0.18$  &  $2.06 \pm 0.18$  \\
SDSSJ1436$-$0000  &  10.34  &  $226 \pm 17$  &  $11.45 \pm 0.08$  &  $11.69 \pm 0.09$  &  $11.39 \pm 0.02$  &  $ 0.63 \pm  0.07$  &  $\phantom{-}0.35 \pm 0.13$  &  $1.88 \pm 0.19$  \\
SDSSJ1443$+$0304  &  \phantom{0}2.64  &  $218 \pm 11$  &  $10.87 \pm 0.06$  &  $11.12 \pm 0.06$  &  $10.67 \pm 0.02$  &  $ 0.49 \pm  0.08$  &  $\phantom{-}0.09 \pm 0.14$  &  $2.31 \pm 0.12$  \\
SDSSJ1451$-$0239  &  \phantom{0}5.83  &  $224 \pm 14$  &  $11.17 \pm 0.07$  &  $11.39 \pm 0.06$  &  $11.00 \pm 0.03$  &  $ 0.52 \pm  0.08$  &  $\phantom{-}0.19 \pm 0.13$  &  $2.24 \pm 0.19$  \\
SDSSJ1525$+$3327  &  14.13  &  $265 \pm 26$  &  $11.78 \pm 0.09$  &  $12.02 \pm 0.09$  &  $11.72 \pm 0.02$  &  $ 0.63 \pm  0.08$  &  $\phantom{-}0.35 \pm 0.14$  &  $1.77 \pm 0.20$  \\
SDSSJ1531$-$0105  &  \phantom{0}7.54  &  $280 \pm 12$  &  $11.43 \pm 0.09$  &  $11.68 \pm 0.09$  &  $11.35 \pm 0.02$  &  $ 0.61 \pm  0.08$  &  $\phantom{-}0.30 \pm 0.14$  &  $2.13 \pm 0.08$  \\
SDSSJ1538$+$5817  &  \phantom{0}3.63  &  $194 \pm 12$  &  $11.03 \pm 0.08$  &  $11.28 \pm 0.08$  &  $10.80 \pm 0.03$  &  $ 0.45 \pm  0.10$  &  $\phantom{-}0.02 \pm 0.18$  &  $1.90 \pm 0.14$  \\
SDSSJ1614$+$4522$\dagger$  &  \phantom{0}8.18  &  $183 \pm 13$  &  $11.21 \pm 0.13$  &  $11.47 \pm 0.12$  &  $11.07 \pm 0.06$  &  $ 0.54 \pm  0.16$  &  $\phantom{-}0.15 \pm 0.27$  &  $2.00 \pm 0.29$  \\
SDSSJ1621$+$3931  &  \phantom{0}8.85  &  $239 \pm 20$  &  $11.45 \pm 0.06$  &  $11.70 \pm 0.07$  &  $11.41 \pm 0.02$  &  $ 0.64 \pm  0.06$  &  $\phantom{-}0.36 \pm 0.10$  &  $1.80 \pm 0.16$  \\
SDSSJ1627$-$0053  &  \phantom{0}6.87  &  $295 \pm 14$  &  $11.45 \pm 0.09$  &  $11.70 \pm 0.09$  &  $11.30 \pm 0.02$  &  $ 0.54 \pm  0.10$  &  $\phantom{-}0.19 \pm 0.17$  &  $2.33 \pm 0.10$  \\
SDSSJ1630$+$4520  &  \phantom{0}7.80  &  $281 \pm 16$  &  $11.61 \pm 0.07$  &  $11.86 \pm 0.07$  &  $11.43 \pm 0.03$  &  $ 0.50 \pm  0.09$  &  $\phantom{-}0.13 \pm 0.15$  &  $1.97 \pm 0.09$  \\
SDSSJ1636$+$4707  &  \phantom{0}5.93  &  $237 \pm 15$  &  $11.38 \pm 0.08$  &  $11.63 \pm 0.08$  &  $11.13 \pm 0.02$  &  $ 0.42 \pm  0.12$  &  $-0.04 \pm 0.21$  &  $2.09 \pm 0.14$  \\
SDSSJ1644$+$2625  &  \phantom{0}4.49  &  $234 \pm 12$  &  $11.18 \pm 0.09$  &  $11.43 \pm 0.08$  &  $11.00 \pm 0.02$  &  $ 0.50 \pm  0.10$  &  $\phantom{-}0.12 \pm 0.16$  &  $2.09 \pm 0.10$  \\
SDSSJ1719$+$2939  &  \phantom{0}4.32  &  $295 \pm 15$  &  $11.22 \pm 0.08$  &  $11.46 \pm 0.08$  &  $11.12 \pm 0.02$  &  $ 0.59 \pm  0.08$  &  $\phantom{-}0.28 \pm 0.14$  &  $2.36 \pm 0.09$  \\
SDSSJ2238$-$0754  &  \phantom{0}5.78  &  $200 \pm 11$  &  $11.20 \pm 0.06$  &  $11.45 \pm 0.06$  &  $11.07 \pm 0.02$  &  $ 0.57 \pm  0.07$  &  $\phantom{-}0.23 \pm 0.12$  &  $1.79 \pm 0.12$  \\
SDSSJ2300$+$0022  &  \phantom{0}6.88  &  $284 \pm 17$  &  $11.40 \pm 0.07$  &  $11.65 \pm 0.07$  &  $11.36 \pm 0.02$  &  $ 0.64 \pm  0.06$  &  $\phantom{-}0.37 \pm 0.10$  &  $2.06 \pm 0.13$  \\
SDSSJ2303$+$1422  &  \phantom{0}9.30  &  $253 \pm 16$  &  $11.47 \pm 0.06$  &  $11.71 \pm 0.06$  &  $11.46 \pm 0.02$  &  $ 0.67 \pm  0.05$  &  $\phantom{-}0.42 \pm 0.09$  &  $1.86 \pm 0.13$  \\
SDSSJ2321$-$0939  &  \phantom{0}6.52  &  $246 \pm \phantom{0}8$  &  $11.35 \pm 0.08$  &  $11.60 \pm 0.08$  &  $11.20 \pm 0.02$  &  $ 0.54 \pm  0.09$  &  $\phantom{-}0.19 \pm 0.16$  &  $2.02 \pm 0.08$  \\
SDSSJ2341$+$0000  &  10.59  &  $206 \pm 13$  &  $11.48 \pm 0.08$  &  $11.73 \pm 0.08$  &  $11.45 \pm 0.02$  &  $ 0.65 \pm  0.07$  &  $\phantom{-}0.38 \pm 0.12$  &  $1.62 \pm 0.12$  \\

%% file: T_ml_fits.tex
L$_B$ ($B$-band)  &  $0.02\pm0.04$ &  $0.59\pm0.03$ &  $0.02\pm0.01$ &&  $0.19\pm0.06$ &  $0.85\pm0.05$ &  $0.09\pm0.01$ \\
L$_B$ ($V$-band)  &  $0.01\pm0.04$ &  $0.49\pm0.03$ &  $0.01\pm0.01$ &&  $0.19\pm0.06$ &  $0.75\pm0.05$ &  $0.09\pm0.01$ \\
L$_V$ ($B$-band)  &  $0.03\pm0.04$ &  $0.58\pm0.04$ &  $0.02\pm0.01$ &&  $0.21\pm0.06$ &  $0.82\pm0.05$ &  $0.09\pm0.01$ \\
L$_V$ ($V$-band)  &  $0.02\pm0.03$ &  $0.49\pm0.03$ &  $0.01\pm0.01$ &&  $0.20\pm0.06$ &  $0.71\pm0.05$ &  $0.09\pm0.01$ \\
$\sigma_{e/2}$ ($B$-band)  &  $0.26\pm0.16$ &  $0.50\pm0.07$ &  $0.02\pm0.01$ &&  $0.89\pm0.23$ &  $0.63\pm0.09$ &  $0.09\pm0.01$ \\
$\sigma_{e/2}$ ($V$-band)  &  $0.19\pm0.14$ &  $0.42\pm0.06$ &  $0.01\pm0.01$ &&  $0.81\pm0.23$ &  $0.56\pm0.09$ &  $0.09\pm0.01$ \\
M$_{\rm dim}$ ($B$-band)  &  $0.03\pm0.03$ &  $0.59\pm0.02$ &  $0.02\pm0.01$ &&  $0.24\pm0.05$ &  $0.83\pm0.03$ &  $0.08\pm0.01$ \\
M$_{\rm dim}$ ($V$-band)  &  $0.02\pm0.03$ &  $0.49\pm0.02$ &  $0.01\pm0.01$ &&  $0.23\pm0.04$ &  $0.72\pm0.03$ &  $0.08\pm0.01$ \\
M$_*$ ($B$-band)  &  $0.07\pm0.04$ &  $0.58\pm0.02$ &  $0.02\pm0.01$ &&  $0.22\pm0.06$ &  $0.91\pm0.02$ &  $0.09\pm0.01$ \\
M$_*$ ($V$-band)  &  $0.05\pm0.03$ &  $0.48\pm0.01$ &  $0.01\pm0.01$ &&  $0.21\pm0.06$ &  $0.81\pm0.02$ &  $0.09\pm0.01$ \\
M$_{r_{\rm e}/2}$ ($B$-band)  &  $0.04\pm0.03$ &  $0.60\pm0.01$ &  $0.02\pm0.01$ &&  $0.27\pm0.04$ &  $0.93\pm0.01$ &  $0.07\pm0.01$ \\
M$_{r_{\rm e}/2}$ ($V$-band)  &  $0.03\pm0.03$ &  $0.50\pm0.01$ &  $0.01\pm0.01$ &&  $0.26\pm0.04$ &  $0.83\pm0.01$ &  $0.07\pm0.01$ \\

%% file: T_dm_fits_2d.tex
L$_B$  &  $0.16\pm0.06$ &  $0.47\pm0.05$ &  $0.05\pm0.02$ &&  $0.28\pm0.10$ &  $\phantom{-}0.08\pm0.08$ &  $0.09\pm0.03$ \\
L$_V$  &  $0.16\pm0.06$ &  $0.46\pm0.05$ &  $0.05\pm0.02$ &&  $0.28\pm0.10$ &  $\phantom{-}0.04\pm0.09$ &  $0.09\pm0.03$ \\
$\sigma_{e/2}$  &  $0.46\pm0.22$ &  $0.40\pm0.09$ &  $0.06\pm0.02$ &&  $0.80\pm0.44$ &  $-0.05\pm0.18$ &  $0.11\pm0.03$ \\
$r_{\rm e}$  &  $0.28\pm0.05$ &  $0.36\pm0.05$ &  $0.03\pm0.02$ &&  $0.49\pm0.10$ &  $-0.13\pm0.09$ &  $0.06\pm0.03$ \\
M$_*$  &  $0.13\pm0.06$ &  $0.54\pm0.03$ &  $0.06\pm0.02$ &&  $0.23\pm0.11$ &  $\phantom{-}0.14\pm0.07$ &  $0.10\pm0.03$ \\
M$_{r_{\rm e}/2}$  &  $0.20\pm0.04$ &  $0.54\pm0.02$ &  $0.03\pm0.02$ &&  $0.36\pm0.07$ &  $\phantom{-}0.20\pm0.03$ &  $0.06\pm0.03$ \\